\documentclass[amsmath,amssymb,aps,showkeys,reprint, %
floatfix,groupedaddress,superscriptaddress,amsmath,amssymb, %
prd]{revtex4-1}

\usepackage{dcolumn}
\usepackage{amsmath,amssymb}
\usepackage[breaklinks = true,colorlinks = true,%
linkcolor = blue,citecolor = blue]{hyperref}
\usepackage{graphicx}
\usepackage{xcolor}
\usepackage{ascmac}
\usepackage{slashed}

\newcommand{\diag}{\mathrm{diag}}

\newcommand{\calC}{\mathcal{C}}
\newcommand{\calF}{\mathcal{F}}

\newcommand{\nB}{n_{\rm B}}

\newcommand{\thetab}{\theta_{\rm B}}

\newcommand{\dd}{\langle dd \rangle}
\newcommand{\phiud}{\Phi_{\rm 2SC}}
\newcommand{\phidd}{\Phi_{dd}}
\newcommand{\deltadd}{\Delta_{dd}}
\newcommand{\deltaud}{\Delta_{\rm 2SC}}
\newcommand{\phiudop}{\hat{\Phi}_{\rm 2SC}}
\newcommand{\phiddop}{\hat{\Phi}_{dd}}
\newcommand{\qop}{\hat{q}}

\usepackage{ulem}

\begin{document}

\title{Non-Abelian Alice strings in two-flavor dense QCD}

\author{Yuki~Fujimoto}
\email{fujimoto@nt.phys.s.u-tokyo.ac.jp}
\affiliation{Department of Physics, The University of Tokyo,
  7-3-1 Hongo, Bunkyo-ku, Tokyo 113-0033, Japan}

\author{Muneto~Nitta}
\email{nitta@phys-h.keio.ac.jp}
\affiliation{Department of Physics \& Research and Education Center for Natural Sciences,
Keio University, Hiyoshi 4-1-1, Yokohama, Kanagawa 223-8521, Japan}

\begin{abstract}

  Quark-hadron continuity with two-flavor quarks that was proposed
  recently connects hadronic matter with neutron $^3P_2$ superfluidity
  and two-flavor dense quark matter.  This two-flavor dense quark
  phase consists of the coexistence of the 2SC condensates and the
  $P$-wave diquark condensates of $d$-quarks, which gives rise to
  color superconductivity as well as superfluidity.
  We classify vortices in this phase.  The most stable vortices are
  what we call the non-Abelian Alice strings, which are 
  superfluid vortices with non-Abelian color magnetic fluxes therein,
  exhibiting so-called topological obstruction, or a non-Abelian
  generalization of the Alice property.
  We show that a single Abelian superfluid vortex is unstable against
  decay into three non-Abelian Alice strings.
  We discover that a non-Abelian Alice string carries orientational
  moduli of the real projective space ${\mathbb R}P^2$ corresponding
  to the color flux therein in the presence of the $P$-wave condensates
  alone.
  We calculate Aharanov-Bohm (AB) phases around the non-Abelian Alice
  string, and find that the 2SC condensates and string's orientational
  moduli must be aligned with each other because of single-valuedness
  of the AB phases of the 2SC condensates.
   
\end{abstract}
\maketitle

\section{Introduction}
Color superconductor is the ground state of the cold QCD matter at
densities much higher than that of saturated nuclei
$n_0=0.16~\text{fm}^{-3}$; the only
known circumstance where we might find such kind of matter is in the
core of neutron stars~\cite{Alford:2007xm}.  Various phases are known for
color superconductivity such as color-flavor locked (CFL)
phase~\cite{Alford:1998mk} in three-flavor symmetric matter and
2-flavor superconducting (2SC) phase~\cite{Alford:1997zt, Rapp:1997zu}
in two-flavor symmetric matter.

Quantum vortices or flux tubes arise in color-superconducting quark
matter~\cite{Eto:2013hoa}. 
In the CFL phase, topologically stable superfluid vortex comes
about owing to the nontrivial first homotopy group
$\pi_1[\mathrm{U(1)_B}] =\mathbb{Z}$~\cite{Forbes:2001gj,
  Iida:2002ev}.  It is related with the broken $\mathrm{U(1)_B}$
symmetry in the CFL phase, which is possible as the Vafa-Witten
theorem does not apply at finite density~\cite{Vafa:1983tf}.
The minimal stable configuration in the CFL phase is known to be the
non-Abelian semi-superfluid vortices carrying color magnetic flux and
only 1/3 of the circulation of the Abelian superfluid
vortices~\cite{Balachandran:2005ev, Nakano:2007dr, Nakano:2008dc,
  Eto:2009kg, Eto:2013hoa}.
Recently, in the context of the \textit{quark-hadron
  continuity}~\cite{Schafer:1998ef, Alford:1999pa, Fukushima:2004bj,
  Hatsuda:2006ps, *Yamamoto:2007ah, Hatsuda:2008is,
  Schmitt:2010pf}---the concept that color superconductor and hadronic
superfluid are continuously connected building upon the identical
symmetry breaking patterns and low-lying excitations in the both
phases---connection of the CFL vortices with the hadronic ones has
been discussed~\cite{Alford:2018mqj, Chatterjee:2018nxe,
  *Chatterjee:2019tbz, Cherman:2018jir, Hirono:2018fjr, Hirono:2019oup,
  Cherman:2020hbe}.

\begin{figure}
    \centering
    \includegraphics[width=\columnwidth]{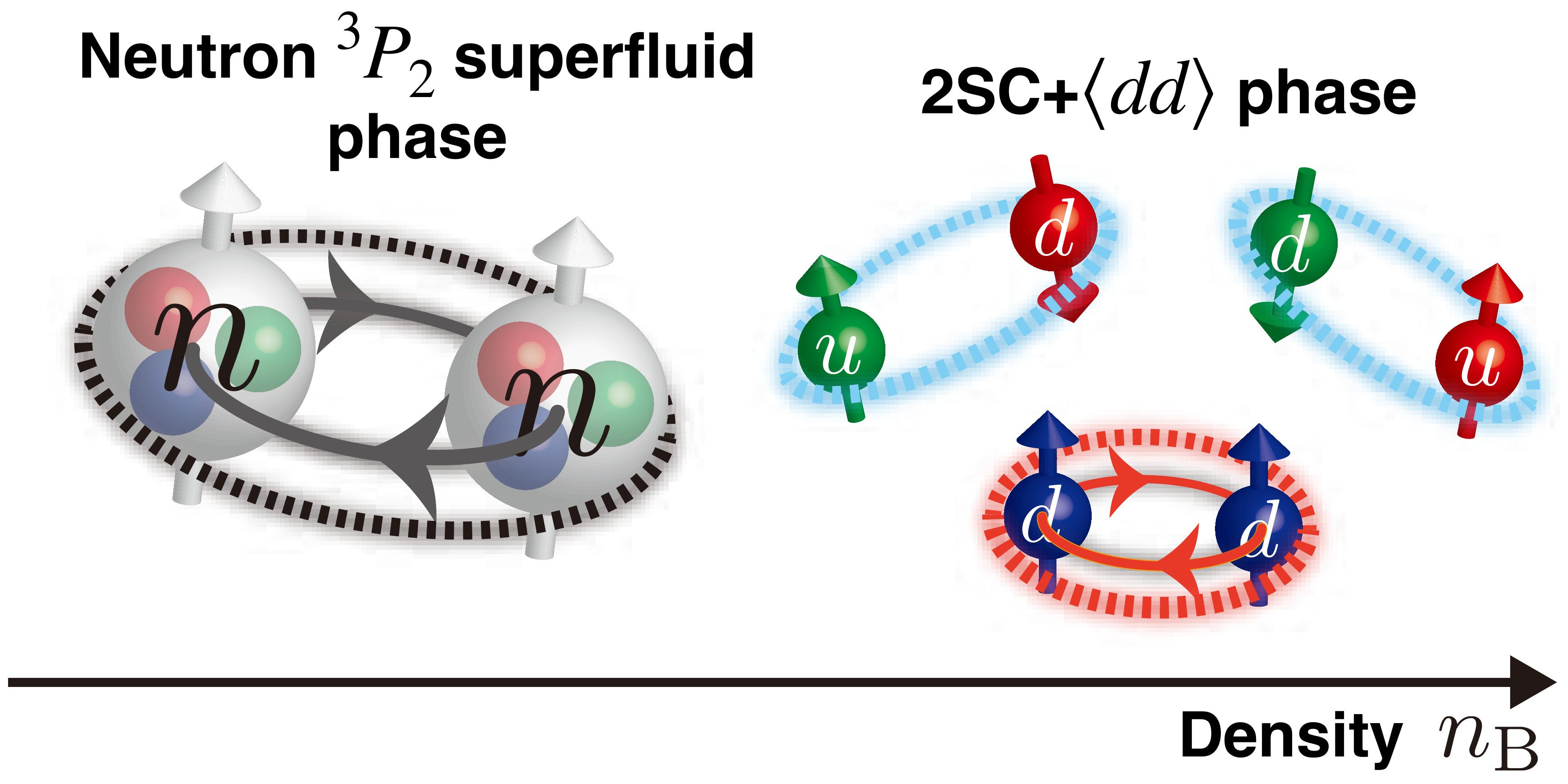}
    \caption{Illustration of the quark-hadron continuity from the
      neutron $^3P_2$ superfluid phase to the 2SC+$\dd$ phase.}
    \label{fig:qhc}
\end{figure}

In contrast, in the 2SC phase, it has been thought that
topologically stable vortices do not appear since the
$\mathrm{U(1)_B}$ stays intact resulting in a trivial first homotopy
group, unlike in the CFL phase~\cite{Alford:2010qf}.
The recent insights from the neutron star observations, however, lead
to the novel phase of color superconductivity in two-flavor matter
with the broken $\mathrm{U(1)_B}$ symmetry~\cite{Fujimoto:2019sxg,
  *Fujimoto:2020cho}.  This phase is called the 2SC+$\dd$ phase.  
Here, we comment on how the 2SC+$\dd$ phase comes into play.
Our current understanding of the cold high-density QCD matter relies
on the quantity called \textit{equation of state} (EoS), which is a
pressure function $P(n_{\rm B})$ of the baryon number density $n_{\rm
  B}$ at $T=0$.
At densities around $n_{\rm B} \sim n_0$, description by nucleon
degrees of freedom works well and the matter is dominated by neutrons.
Meanwhile, at higher densities, the matter is well described in terms
of quark degrees of freedom.
According to the current model-independent analysis of the neutron
star EoS, there might be a possibility that a substantial quark core
exists inside a heavy neutron star~\cite{Annala:2019puf}.
The sizable quark core inside the neutron star can be realized by a
smooth crossover from the hadronic matter to the quark matter in the
EoS~\cite{Masuda:2012kf,*Masuda:2012ed, Kojo:2014rca, Baym:2017whm}.
This crossover construction satisfies the stringent constraints
put by observations, e.g., the two-solar-mass
conditions~\cite{Demorest:2010bx,
  *Fonseca:2016tux,*Antoniadis:2013pzd, *Cromartie:2019kug}.
The rationale behind the crossover construction is the concept of
quark-hadron continuity.
The quark-hadron continuity has been previously formulated in the
ideal $\mathrm{SU(3)_F}$ flavor symmetric setup~\cite{Schafer:1998ef,
  Alford:1999pa, Fukushima:2004bj, Hatsuda:2006ps, *Yamamoto:2007ah,
  Hatsuda:2008is}.
At $\nB \sim n_0$, the matter is dominated by neutrons, which are
composed of two-flavor $u$ and $d$ valence quarks, not by hyperons
that contains $s$ quark.
Moreover, neutrons are known to be paired up in the $^3P_2$ channel and
show the superfluidity~\cite{Hoffberg:1970vqj, Tamagaki:1970ptp,
  *takatsukaPTP71, *takatsukaPTP72, richardsonPRD72, Sauls:1978lna,
  Takatsuka:1992ga} (see also Refs.~\cite{Mizushima:2016fbn,
  Yasui:2018tcr, Yasui:2019unp} and references therein for recent
studies).
To fit with this natural ground state at $\nB \sim n_0$, we are lead to
the more realistic picture of quark-hadron continuity in the
$\mathrm{SU(2)_F}$ flavor symmetric setup---the neutron $^3 P_2$
superfluid is continuously connected to the two-flavor quark
matter---in accordance with the above-mentioned crossover EoS.
As a consequence, the new condensate $\langle \hat{d}^T \calC \gamma^i
\nabla^j \hat{d} \rangle$, which is a diquark of $d$-quarks paired in
the $P$-wave channel, appears in addition to the conventional two-flavor
2SC condensate.  We will use the shorthand notation $\dd$ instead of
$\langle \hat{d}^T \calC \gamma^i \nabla^j \hat{d} \rangle$ throughout
this work unless otherwise specified.  The coexistence of the 2SC and
$\dd$ condensates in the 2SC+$\dd$ phase ensures the continuity to
hold as illustrated in Fig.~\ref{fig:qhc}.

In this work, on the basis of this 2SC+$\dd$ phase picture, we study
vortices arising in the two-flavor dense quark matter for the first
time.  We describe that the most stable vortices in the $\dd$
phase are non-Abelian vortices, which support 1/3 fractional windings
in $\rm U(1)_B$ as well as the color-magnetic fluxes similar to those
in the CFL phase.  These vortices also exhibit the unique features
akin to the so-called Alice strings~\cite{Schwarz:1982ec,
  Alford:1990mk, Alford:1990ur, Alford:1992yx, Preskill:1990bm, 
  Bucher:1992bd, Lo:1993hp, Leonhardt:2000km,
  Chatterjee:2017jsi,
  Chatterjee:2017hya, Chatterjee:2019zwx}, thus we named them as
``non-Abelian Alice strings''.
This feature is what is called topological obstruction, implying that
some unbroken generators in the bulk are not globally defined around
the string.
We also show that a single Abelian $\rm U(1)_B$ vortex is unstable
against a decay into triple of non-Abelian Alice strings, as the case
of those in the CFL phase~\cite{Nakano:2007dr,Cipriani:2012hr,
Alford:2016dco, Chatterjee:2018nxe, *Chatterjee:2019tbz}.
We then show that a non-Abelian Alice string carries orientational
moduli, or collective coordinates, of the real projective space ${\mathbb R}P^2$ corresponding to
the color flux therein in the presence of the $\dd$ condensates alone,
in contrast to those in the CFL phase carrying moduli of the complex
projective space ${\mathbb
  C}P^2$~\cite{Nakano:2007dr,Eto:2009bh,Eto:2013hoa,Eto:2009tr}.  We
show that quasi-quarks exhibit nontrivial (generalized) Aharanov-Bohm
(AB) phases around the non-Abelian Alice string, and then find the
``bulk-soliton moduli locking'' phenomenon, i.e., when the 2SC
condensates develop VEVs, they must be aligned to string's
orientational moduli because of single-valuedness of the AB phases of
them.

This paper is organized as follows.
In Sec.~\ref{sec:nf2}, we will further review the structure of the
2SC+$\dd$ phase.  We will consider the corresponding symmetry breaking
patterns.
In Sec.~\ref{sec:minimal}, we will categorize vortices in the presence
of $\dd$ diquark into three-types: Abelian superfluid vortices,
color-magnetic flux tubes, and the non-Abelian Alice strings.
In Sec.~\ref{sec:properties}, we will discuss properties of the
non-Abelian Alice string.  We will also introduce the generalized
AB phase around strings, and discuss that the Abelian
vortices break up into three Alice strings based on the AB phase.
The most stable configuration among these turns out to be the
non-Abelian Alice string.
In Sec.~\ref{sec:bulksoliton}, we will point out that the moduli of
vortex are aligned with the bulk quantity in the point of view of
generalized AB phase.
We will make a comment on the consistency on the ordering of the
diquark condensate formation in Sec.~\ref{sec:consistency}.  In
Sec.~\ref{sec:summary}, we will finally summarize our results.

\section{Two-flavor dense quark matter}
\label{sec:nf2}

In this section, after giving a summary of the two-flavor dense quark matter,
especially the 2SC+$\dd$ phase 
formulated in Refs.~\cite{Fujimoto:2019sxg, *Fujimoto:2020cho}, 
we discuss symmetry breaking patterns in this phase.

\subsection{2SC+$\dd$ phase from quark-hadron continuity}

The concept of quark-hadron continuity builds upon the so-called
Fradkin-Shenker theorem~\cite{Osterwalder:1977pc, Fradkin:1978dv,
  Banks:1979fi} (see also~\cite{Cherman:2020hbe} for recent model
study on the validity of this theorem).
This folk theorem states that within a gauge-Higgs theory, we cannot
distinguish between the confinement and Higgs phases with 
any local order parameters 
as long as global symmetries in these phases 
are the same.
We apply this theorem to the two-flavor dense QCD, 
where confinement and Higgs phases correspond to 
the hadronic and color-superconducting quark phases, respectively.
Here, we take a superfluid operator as an example of a local
order parameter with which we cannot distinguish these two phases.

The hadronic phase in this case is 
a neutron $^3 P_2$ superfluid, 
for which the order parameter operator is given
by~\cite{Hoffberg:1970vqj, Tamagaki:1970ptp, Takatsuka:1992ga}
\begin{align}
  \hat{A}^{ij} &= \hat{n}^T \calC \gamma^i \nabla^j \hat{n}\,,
\end{align}
where $\hat{n}$ denotes a neutron field operator,
$\calC$ is the charge conjugation operator, 
and indices ($i,j,\ldots$) denote spatial coordinates.  
In this paring, the matrices $\gamma^i$ and spatial derivatives 
$\nabla^j$ account for spin and angular momentum
contributions in the $^3 P_2$ pairing, respectively. 
We further assume that neutrons made out of $u$ and $d$-quarks can be
described as a quark-diquark system $\hat{n} = \epsilon^{\alpha\beta\gamma}
(\hat{u}_\alpha^T \calC \gamma^5 \hat{d}_\beta) \hat{d}_\gamma$ with 
the Greek letters ($\alpha,\beta,\ldots$) being the color indices.
With this in mind, $\hat{A}^{ij}$ can be rearranged into three diquarks as
\begin{align}
  \hat{A}^{ij} &\propto
  \epsilon^{\alpha \beta \gamma} \epsilon^{\alpha' \beta' \gamma'}
  (\hat{u}^T_\alpha \calC \gamma^5 \hat{d}_{\beta})
  (\hat{u}^T_{\alpha'} \calC \gamma^5 \hat{d}_{\beta'})
  (\hat{d}^T_\gamma \calC \gamma^i \nabla^j \hat{d}_{\gamma'})\notag\\
  &=(\phiudop)^\gamma(\phiudop)^{\gamma'}(\phiddop)^{ij}_{\gamma\gamma'}
  \label{eq:factorize}
\end{align}
with $\phiudop$ and $\phiddop$ being the diquark operators defined as
\begin{align}
  (\phiudop)^\alpha &\equiv \epsilon^{\alpha\beta\gamma} 
  \hat{u}^T_{\beta} \calC \gamma^5 \hat{d}_{\gamma} \,, \label{eq:2scop}\\
  (\phiddop)_{\alpha\beta}^{ij} &\equiv  \hat{d}^T_{\alpha} \calC
  \gamma^i \nabla^j \hat{d}_\beta\,. \label{eq:ddop}
\end{align}

One can take the expectation value of $\hat{A}^{ij}$ in the hadronic phase,
which reads 
\begin{align}
  \langle \hat{A}^{ij} \rangle &= \langle \hat{n}^T \calC \gamma^i \nabla^j \hat{n}
  \rangle \,.
\end{align}
This condensate accounts for the $^3P_2$ superfluidity of the neutron
matter.

By contrast, one can also take the expectation value of $\hat{A}^{ij}$ in
the quark phase under the mean field approximation in light of
Eq.~\eqref{eq:factorize}:
\begin{align}
  \langle \hat{A}^{ij} \rangle \simeq
  (\phiud)^\alpha (\phiud)^\beta
  (\phidd)_{\alpha\beta}^{ij}\,,
  \label{eq:aij}
\end{align}
where we have defined
\begin{align}
  \phiud &\equiv \langle \phiudop \rangle\,, \label{eq:2sc}\\
  \phidd &\equiv \langle \phiddop \rangle\,. \label{eq:dd}
\end{align}
Note that quantities without hat symbols denote condensates while those
with hat symbols are operators.
The two condensates $\phiud$ and $\phidd$ at the mean field
level in Eq.~\eqref{eq:aij} account for the color superconductivity of
the quark matter:  
$\phiud$ is the so-called 2SC condensate, 
while 
the novel feature here is represented by $\phidd$, 
which is the diquark
condensate of $d$-quarks in the $^3P_2$ channel.  We understand the indices
$\alpha, \beta$ of $(\phidd)_{\alpha\beta}$ are in $\boldsymbol{6}$
representation of color as they should be symmetric.
Qualitatively speaking, this $^3P_2$ pairing arises owing to the
short range repulsion in the color $\boldsymbol{6}$ channel of
one-gluon exchange (OGE), which disfavors the $S$-wave pairing,
and the selective attraction in the $J=2$ channel, which arises from
the spin-orbit dependent part of the Fermi-Breit reduced interaction
of OGE.
We call this coexistence phase of $\phiud$ and $\phidd$ as 
the \textit{2SC+$\dd$ phase}.

As mentioned above, $\langle A^{ij} \rangle$ is always nonzero in
both phases, so the local order parameter indeed cannot distinguish
these two phases.  It leads us to the continuity.
For more details, see Ref.~\cite{Fujimoto:2019sxg, *Fujimoto:2020cho}.

\subsection{Symmetry of the color-superconducting phase}

We consider the symmetry group of QCD as $G_{\rm QCD} =
\mathrm{SU(3)}_{\rm C} \times \mathrm{U(1)}_{\rm B}$, 
without paying attention to the chiral symmetry of QCD 
since it does not play an important role in discussion of this paper.
Under the element $(U, e^{i \thetab}) \in \mathrm{SU(3)_C
  \times U(1)_B}$  acting on quark fields $q$, as a column vector
belonging to the fundamental representation $\boldsymbol{3}$ of
$\mathrm{SU(3)_C}$, as
\begin{equation}
  q \to  e^{i \thetab} U q,
\end{equation}
where $U \in$ (the $\boldsymbol{3}$ representation of)
$\mathrm{SU(3)_C}$, the diquark
condensates~(\ref{eq:2sc},~\ref{eq:dd}) transforms as
\begin{equation}
  \phiud \to e^{2i\theta_{\rm B}} U^{\ast} \phiud\,,\quad
  \phidd \to e^{2i\theta_{\rm B}} U \phidd U^T\,.
  \label{eq:transf}
\end{equation}
Here, the factor 2 in $\mathrm{U(1)_B}$
is inserted to maintain the consistency with a single quark;
we suppress the spatial indices $i,j$ of $\phidd$ hereafter,
where an appropriate tensor structure is implied~\footnote{It is known
  in the nematic phase~\cite{Sauls:1978lna} for which
  $\diag(1,r,1-r)$ is implied for the $i,j$ indices, with real
  parameter $r$.}.

\begin{figure}[t]
    \centering
    \includegraphics[width=\columnwidth]{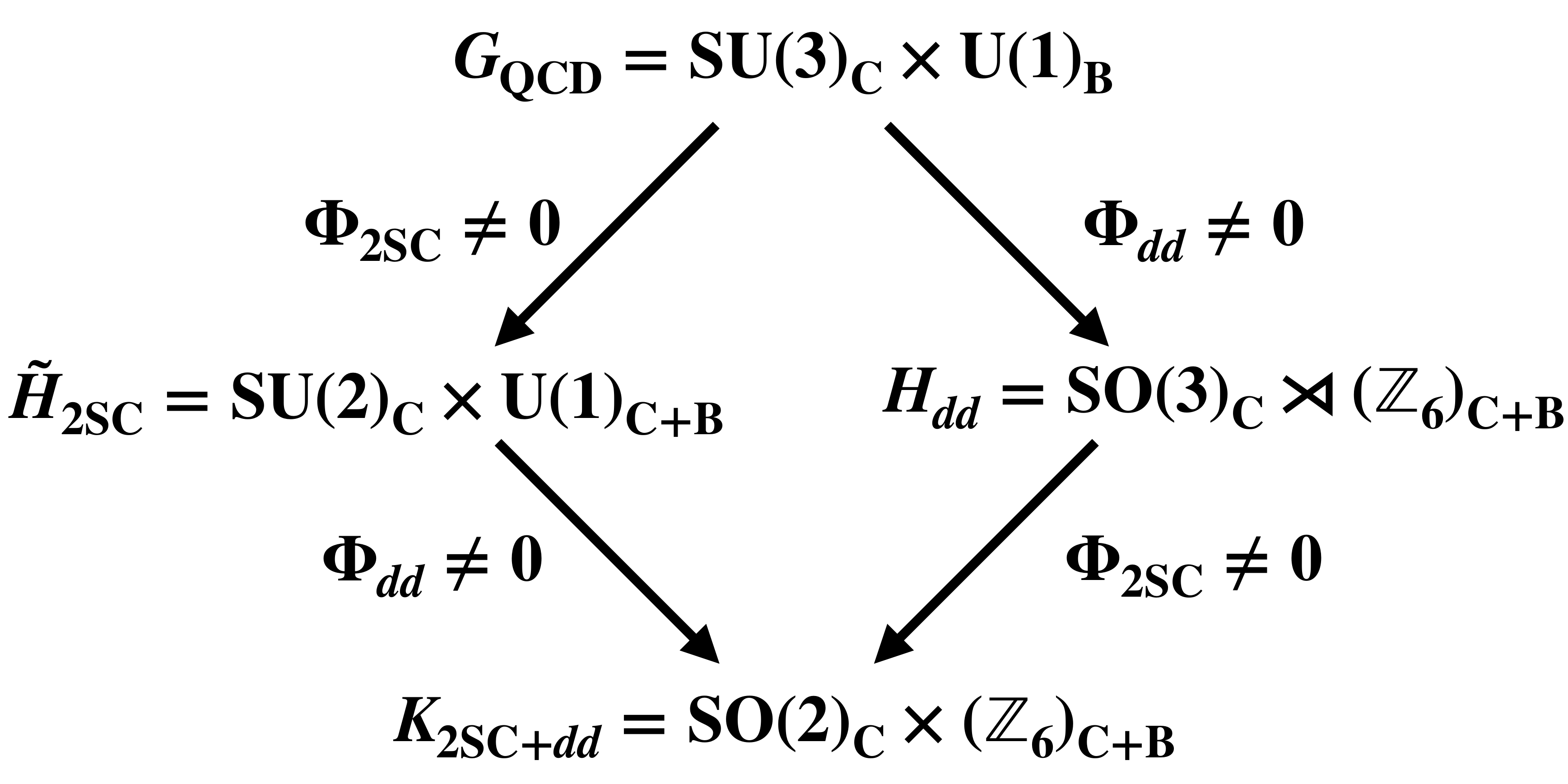}
    \caption{Spontaneous symmetry breaking patterns 
    in the most symmetric case.}
    \label{fig:SSB}
\end{figure}

Once the condensates $\phiud$ and $\phidd$ are nonzero,
symmetry is spontaneously broken and superconductivity sets in.  
Here, we examine the subgroups that leave these condensates invariant
and identify the corresponding symmetry breaking patterns.
We consider two regimes in which each condensate is turned on
sequentially instead of turning them on simultaneously.  Namely, we
firstly turn on $\phidd$ and then turn on $\phiud$ and choose the
gauge so as to be consistent with each other, and vice versa.
This can be summarized as
\begin{align}
  1. \quad & G_{\rm QCD} \xrightarrow{\phidd} H_{dd}
  \xrightarrow{\phiud} K_{\mathrm{2SC}+dd} \label{eq:ghk} \\
  2. \quad & G_{\rm QCD} \xrightarrow{\phiud} \tilde{H}_{\rm 2SC}
  \xrightarrow{\phidd} K_{\mathrm{2SC}+dd} \label{eq:ghtildek}
\end{align}
and also in Fig.~\ref{fig:SSB}.  We will mainly focus on the first
option in this work.  The second option is also mentioned to ensure
the consistency.

\subsection{Symmetry breaking by the $\dd$ condensate}

First, let us turn to $G_{\rm QCD} \xrightarrow{\phidd} H_{dd}$
part in Eq.~\eqref{eq:ghk}.
The matrix $(\phidd)_{\alpha\beta}$ can be diagonalized by a gauge
rotation, so that without loss of generality one can take
\begin{equation}
  \phidd = \diag\left[(\phidd)_{11}, (\phidd)_{22}, (\phidd)_{33}\right]\,.
\end{equation}
Here, for simplicity, 
we posit the most symmetric case 
that in the ground state $(\phidd)_{11} = (\phidd)_{22}
= (\phidd)_{33} = \deltadd$, thus 
\begin{equation}
\phidd = \deltadd
\boldsymbol{1}_3.
\end{equation}
Then, the unbroken subgroup
\begin{equation}
  H_{dd} = \mathrm{SO(3)_C}\rtimes (\mathbb{Z}_6)_{\rm C+B}
  \label{eq:Hdd}
\end{equation}
keeps $\phidd = \deltadd \boldsymbol{1}_3$ invariant.
The condition $UU^T = \boldsymbol{1}_3$, which comes
from Eq.~\eqref{eq:transf} taking $\thetab=0$, imposes
$\mathrm{SO(3)_C}$.
The discrete group is defined by
\begin{equation}
  (\mathbb{Z}_6)_{\rm C+B} : (X^k, \omega^{-2k}) \in \mathrm{SU(3)_C
    \times U(1)_B}
  \label{eq:z6}
\end{equation}
with $k=0,1,2,3,4,5$ and $\omega \equiv e^{i\pi/3}$ being 
the sixth root of unity;
\begin{equation}
 X \equiv \diag(\omega, \omega, \omega^{-2}) \label{eq:X1}
\end{equation} 
 is generated by $T_8
\propto \diag(1,1,-2)$ of the broken $\mathrm{SU(3)_C}$ symmetry.
We define the semidirect product $\rtimes_\phi$ in $H_{dd}$ by the
group operation $\circ$ determined by the homomorphism $\phi$:
\begin{align}
  (U, X^k) \circ (U', X^{k'}) &= (U\phi_{X^k}(U'),
  X^k X^{k'})\,, \\
  \phi_{X^k}(U') &\equiv X^k U' (X^k)^{-1}\,,
\end{align}
where $(U^{(\prime)}, X^{k^{(\prime)}}) \in \mathrm{SO(3)_C}
\times (\mathbb{Z}_6)_{\rm C+B}$.  $\phi_{X^k}$ is trivial when $k$ is
even, but is nontrivial when $k$ is odd.
Note that the explicit definition of $(\mathbb{Z}_6)_{\rm C+B}$ in
Eq.~\eqref{eq:z6} is not unique due to the semidirect product in
$H_{dd}$, so one can also take, e.g., 
\begin{equation}
 X=\diag(\omega^{-2}, \omega,
\omega)  \mbox{ or }  
X=\diag(\omega, \omega^{-2}, \omega). \label{eq:X23}
\end{equation}

The order parameter manifold for the symmetry breaking $G_{\rm QCD}
\to H_{dd}$ by $\phidd$ is
\begin{equation}
  \frac{G_{\rm QCD}}{H_{dd}} 
  = \frac{\mathrm{SU(3)_{C}} \times \mathrm{U(1)_B}}{\mathrm{SO(3)_C} \rtimes (\mathbb{Z}_6)_{\rm C+B}} 
  \simeq
  \frac{M_3 \times S^1}{(\mathbb{Z}_6)_{\rm C+B}}
  \label{eq:G/H}
\end{equation}
with $M_3 \equiv \mathrm{SU(3)_C}/\mathrm{SO(3)_C}$.
The ground state allows topologically stable vortex configurations since
$\pi_1(G_{\rm QCD}/H_{dd}) = \mathbb{Z}$. See
Refs.~\cite{Auzzi:2006ns, Auzzi:2008hu} for homotopy groups of $M_3$.
The broken generators of the coset space $M_3$
belong to ${\bf 5}$ representation  (traceless
symmetric $2 \times 2$ tensor) of SO(3)$_{\rm C}$.

\subsection{Symmetry breaking by the 2SC condensate}
\label{sec:SSB2SC}

Next, we turn on the $\phiud$ in the presence of
$\phidd$ and address $H_{dd} \xrightarrow{\phiud}
K_{\mathrm{2SC}+dd}$ part in Eq.~\eqref{eq:ghk}.
We can divide states into two cases according to the real or complex
nature of $\phiud$.  We will only focus on the case $\Phi_{\rm 2SC}
\in \mathbb{R}^3$ in this paper;  The other possibility $\phiud \in
\mathbb{C}^3$, which also shows interesting physics, will be discussed
in the forthcoming paper.
In the case of $\Phi_{\rm 2SC} \in \mathbb{R}^3$, 
$(\phiud)^\alpha$ can be transformed to one component by the
unbroken gauge symmetry $\mathrm{SO(3)_{\rm C}}$ in $H_{dd}$ in
Eq.~\eqref{eq:Hdd}, so that without loss of generality we can write
\begin{equation}
  (\phiud)^\alpha = \deltaud \delta^{\alpha3}\,.\label{eq:2SC-R3}
\end{equation}

Let us consider the $(\mathbb{Z}_6)_{\rm C+B}$ action on $\Phi_{\rm
  2SC}$ in Eq.~\eqref{eq:transf}.
\begin{align}
  &\phiud \to \omega^{-2} X^{-1} \phiud \notag\\
  &= \left\{\begin{array}{c}
  \omega^{-2} \,\diag(\omega, \omega, \omega^{-2})^{-1}\phiud = \diag(-1, -1, 1) \phiud \\
  \omega^{-2} \,\diag(\omega, \omega^{-2}, \omega)^{-1}\phiud = \diag(-1, 1, -1) \phiud \\
  \omega^{-2} \,\diag(\omega^{-2}, \omega, \omega)^{-1}\phiud = \diag(1, -1, -1) \phiud .\\
  \end{array}\right. 
  \label{eq:2SCZ6}
\end{align}
Then, the direction of $X \in \mathbb{Z}_6$ inside $G_{\rm QCD}$, 
namely Eq.~(\ref{eq:X1}) or (\ref{eq:X23}),
is determined according to
the value of $\phiud$.  Now we took $(\phiud)^\alpha =
\deltaud \delta^{\alpha3}$, so the symmetry
$(\mathbb{Z}_6)_{\rm C+B}$ in Eq.~\eqref{eq:2SCZ6} remains unbroken
because of the first line in Eq.~\eqref{eq:2SCZ6}.  Eventually, the 
nonzero component in $\phiud$ fixes where the $\omega^{-2}$ appears
in the $X \in \mathbb{Z}_6$ transformation~\eqref{eq:z6} 
with Eq.~\eqref{eq:X1}.  An SO(2)$_{\rm C}$ group remains unbroken,
and the semidirect product is reduced to the direct product with the
unbroken SO(2)$_{\rm C}$ group:
\begin{align}
  H_{dd} \to K_{\mathrm{2SC}+dd} =
  \mathrm{SO(2)}_{\rm C} \times (\mathbb{Z}_6)_{\rm C+B}.
  \label{eq:HtoK}
\end{align}
Thus, the order parameter manifold for this symmetry breaking is 
\begin{align}
  \frac{H_{dd}}{K_{\mathrm{2SC}+dd}} 
  =
  \frac{\mathrm{SO(3)_C} \rtimes (\mathbb{Z}_6)_{\rm C+B}}{\mathrm{SO(2)}_{\rm C} \times (\mathbb{Z}_6)_{\rm C+B}}
  \simeq \frac{\mathrm{SO(3)_C}}{\mathrm{SO(2)}_{\rm C}}
  \simeq S^2\,.
\end{align}

\subsection{The opposite ordering}
\label{sec:opposite}

We turn to the opposite ordering of the condensate formations given in
Eq.~\eqref{eq:ghtildek}.
For the $\phiud$ it is standard to fix the gauge as
$(\phiud)^\alpha = \deltaud \delta^{\alpha3}$, and it
is kept invariant under the symmetry
\begin{equation}
  \tilde{H}_{\rm 2SC} = \mathrm{SU(2)_C \times U(1)_{C+B}}\,.
  \label{eq:htilde}
\end{equation}
The order parameter manifold of this breaking is 
\begin{equation}
    \frac{G_{\rm QCD}}{\tilde{H}_{\rm 2SC}} 
    = \frac{\mathrm{SU(3)_C}}{\mathrm{SU(2)_C}}\,,
  \label{eq:G/H2SC}
\end{equation}
allowing trivial first homotopy group or no stable vortices.

Now we turn on $\phidd$.
The upper-left block of $\phidd$ can be diagonalized by the
$\mathrm{SU(2)_C}$ transformation in $\tilde{H}_{\rm 2SC}$:
\begin{align}
 \phidd =
 \begin{pmatrix}
      \deltadd & 0 & (\phidd)_{13} \\
      0 & \deltadd' & (\phidd)_{23} \\
      (\phidd)_{13} & (\phidd)_{23} & \deltadd''
    \end{pmatrix}\,.\label{eq:gauge-M}
\end{align}
Here, as before, we have assumed the diagonal components to be equal.
To identify an unbroken symmetry,
we see that the ${\mathbb Z}_6$ action in Eq.~\eqref{eq:z6} 
with $X$ in Eq.~(\ref{eq:X1}) acts on $\phidd$ as
\begin{align}
  \phidd 
  &\to
  \omega^{-2} X \phidd X^T \nonumber \\
  &= \begin{pmatrix}
      \deltadd & 0 & \omega^3 (\phidd)_{13} \\
      0 & \deltadd' & \omega^3 (\phidd)_{23} \\
      \omega^3 (\phidd)_{13} & \omega^3 (\phidd)_{23} & \deltadd''
    \end{pmatrix}\,.
\end{align} 
Then, there are two phases depending on the absence or presence of the
off-diagonal components.
When the off-diagonal components vanish,  the unbroken group is $K =
K_{\mathrm{2SC}+dd} \simeq \mathrm{SO(2)_C} \times (\mathbb{Z}_6)_{\rm
  C+B}$ as the same as Eq.~\eqref{eq:HtoK}.
The case that the off-diagonal components are present 
corresponds to the case that 
$\Phi_{\rm
  2SC} \in \mathbb{C}^3$ above Eq.~(\ref{eq:2SC-R3}) 
  for the symmetry breaking in Eq.~(\ref{eq:ghk}).
This case will be
mentioned in the forthcoming paper.

The order parameter manifolds for these breakings are 
\begin{align}
  \frac{\tilde{H}_{\rm 2SC}}{K_{\mathrm{2SC}+dd}} =
  \dfrac{\mathrm{SU(2)_C}}{\mathrm{SO(2)_C}} \times \dfrac{\rm
    U(1)_{C+B}}{(\mathbb{Z}_6)_{\rm C+B}} \simeq S^2 \times
  \dfrac{\rm U(1)_{C+B}}{(\mathbb{Z}_6)_{\rm C+B}}\,.
  \label{eq:OPMH/Kdeconf}
\end{align}
Topologically stable vortex configurations appears in the ground state
according to $\pi_1(\tilde{H}_{\rm 2SC}/K_{\mathrm{2SC}+dd}) = \mathbb{Z}$.

\section{Minimal topological vortex in $\dd$ phase--Non-Abelian Alice string}
\label{sec:minimal}

In this section, we examine the vortices that appear in the
presence of $\phidd$.  As introduced in Sec.~\ref{sec:alice}, we
discuss that the topologically stable minimal configuration is 
a 
non-Abelian vortex rather than a superfluid vortex
(Sec.~\ref{sec:abelian}) or a color-magnetic flux tube
(Sec.~\ref{sec:fluxtube}) similar to the CFL case.
We also discuss that our non-Abelian vortices which show
 the property akin to
that of Alice string, namely, the topological obstruction.

\subsection{Abelian vortex}
\label{sec:abelian}
The simplest vortex is a superfluid vortex of the form
\begin{align}
  \phidd(\varphi) = f_0(r) e^{i \varphi} \deltadd{\bf 1}_3 \sim e^{i
    \varphi} \deltadd{\bf 1}_3\,
  \label{eq:Abelian}
\end{align}
with the boundary condition $f_0(0)=0$ and $f_0(\infty)=1$ 
for the profile function $f$, 
and
$(r, \varphi)$ being the polar coordinates.
This carries a unit quantized circulation in $\mathrm{U(1)_B}$ as
encoded in the factor $e^{i\varphi}$.
We call this a $\rm U(1)_B$ superfluid vortex or an Abelian vortex.
See the first line of Table~\ref{tab:abphase0}.
This string is created under rotation because of the superfluidity.
Although this string is topologically stable due to
$\pi_1[\mathrm{U(1)_B}] = \mathbb{Z}$, it is unstable against decay
into three non-Abelian Alice strings, as discussed below in
Sec.~\ref{sec:decay}.

\subsection{Pure color flux tubes}
\label{sec:fluxtube}

A color-magnetic flux tube generated by a closed loop in the group
manifold $\mathrm{SU(3)_C}$ is given 
(see Refs.~\cite{Iida:2004if,Alford:2010qf} for 
the case of the CFL and 2SC phases, respectively)
\begin{align}
  \begin{split}
    \phidd (\varphi)
    &= \deltadd
    \begin{pmatrix}
      f(r) e^{-2i\varphi}& 0 & 0 \\ 
      0 & f(r) e^{-2i\varphi}& 0 \\ 
      0 & 0 & f(r) e^{4i\varphi}
    \end{pmatrix}\,,\\
    A_i &= - \frac{a(r)}{ g}\frac{\epsilon_{ij} x^j}{r^2} 
    \diag(-1,-1,2),
  \end{split}
  \label{eq:pure-color-1}
\end{align}
where $g$ is the SU(3) gauge coupling constant and the boundary conditions for
the profile functions $f$ and $a$ are set by
\begin{align}
  f (0) =  a(0) = 0, \quad f (\infty)  = a(\infty) = 1\,.
  \label{eq:bc-f}
\end{align}
Here, we have set the condensate at $\varphi = 0$ and large distance as
\begin{align}
  \phidd (\varphi = 0) 
  = \deltadd f(r) \boldsymbol{1}_3\ \label{eq:M(0)}
  \sim \deltadd \boldsymbol{1}_3\,.
\end{align}
This carries a color-magnetic flux
\begin{align}
  \int d^2x F_{12} = \frac{2\pi}{g} \diag(-1,-1,2) =\calF_0 \diag(-1,-1,2)\,,
\end{align}
where $F_{12}$ is the color-magnetic component of the field strength
tensor and we have defined a unit color-magnetic flux $\calF_0$ as
\begin{equation}
  \calF_0 \equiv \frac{2\pi}{g}\,.
  \label{eq:flux}
\end{equation}
This flux tube is unstable and thus decay into the ground state
because of $\pi_1[{\rm SU(3)_C}]=0$.  We will not discuss the color
flux tubes henceforth, and only focus on the topologically stable
objects.

\subsection{Non-Abelian Alice strings}
\label{sec:alice}
Here, we present the most stable vortex in this system, 
that is 
a vortex with orientational moduli in the internal
non-Abelian gauge space, which we call as a non-Abelian Alice string.
This string can be characterized by the holonomy at infinite distance 
\begin{equation}
  U(\varphi) = \mathcal{P} \exp\left(ig\int_0^\varphi
  \boldsymbol{A}\cdot d \boldsymbol{\ell}\right)\,.
  \label{eq:holonomy}
\end{equation}
This generates the condensate winding at spatial infinity:
\begin{equation}
  \phidd(\varphi) = e^{i\varphi/3} U(\varphi) \phidd(\varphi=0) U^T(\varphi)\,.
\end{equation}
The ansatz that we impose here is of the form of
\begin{equation}
  \begin{split}
  \phidd (\varphi)
  &= \deltadd
  \begin{pmatrix}
    g(r)  & 0 & 0 \\ 0 & g(r) & 0 \\ 0 & 0 & f(r) e^{i\varphi}
  \end{pmatrix}\,,\\
  A_i &= - \frac{a(r)}{6 g}\frac{\epsilon_{ij} x^j}{r^2} 
  \diag(-1,-1,2)\,,
  \end{split}
  \label{eq:ansatz}
\end{equation}
where the corresponding holonomy~\eqref{eq:holonomy} is given by
\begin{equation}
    U(\varphi) = e^{i(\varphi/6)\diag(-1,-1,2)} \,,
\end{equation}
and we have set the condensate at $\varphi = 0$ and large distance as
\begin{align}
  \phidd (\varphi = 0) 
  = \deltadd  \begin{pmatrix}
    g(r)  & 0 & 0 \\ 0 & g(r) & 0 \\ 0 & 0 & f(r) 
  \end{pmatrix}
  \sim \deltadd \boldsymbol{1}_3\,.
\end{align}
The boundary conditions for the profiles $f$ and $g$ are
\begin{align}
  f (0) = g' (0) =a(0) = 0, \quad f (\infty) =g(\infty) = a(\infty) = 1.
  \label{eq:bc-fga}
\end{align}
The form of the condensate and the gauge field in
Eq.~\eqref{eq:ansatz} can be understood as follows.  Coupling between
the condensate and the gauge field comes from the minimal coupling in
the covariant derivative $D_\mu \Phi=\partial_\mu \Phi - igA_\mu \Phi$.  The
profile of the vortex can be made such that it only appear in $T_8
\propto \diag(-1,-1,2)$ component of the gauge field with suitable
gauge transformation while keeping the kinetic term $|D_\mu \Phi|^2$
in the energy minimized.
One can factor out the winding in $\mathrm{U(1)_B}$ in
Eq.~\eqref{eq:ansatz} to make the color-magnetic flux clearer
\begin{align}
  \phidd (\varphi)
  &= \deltadd e^{i\varphi/3}
  \begin{pmatrix}
    g(r) e^{-i\varphi/3}  & 0 & 0 \\ 0 & g(r) e^{-i\varphi/3} & 0 \\ 0 & 0 & f(r) e^{2i\varphi/3}
  \end{pmatrix}\,.
\end{align}
From this expression it is evident that this carries $1/6$ quantized
color-magnetic flux,
$\calF = \calF_0 / 6$ (see Eq.~\eqref{eq:flux} for the definition of
$\calF_0$), and $1/3$ quantized circulation in $\rm U(1)_B$, 
as summarized in the second line of Table~\ref{tab:abphase0}.
Since $\phidd$ ought to be single valued, $U(2\pi)$ belongs to the
little group $H_{dd}$ of the condensate $\phidd(0)$.
Therefore, this configuration indeed connects two elements of $\rm
SU(3)_C$: $U(\varphi=0) = {\bf 1}_3$ and $U(\varphi=2\pi) =
e^{i(\pi/3)\diag(-1,-1,2)} = \diag
(\omega^{-1},\omega^{-1},\omega^2)$. Note that the latter does not
belong to the center of $\rm SU(3)_C$.

The configuration in Eq.~(\ref{eq:ansatz}) is not the unique.
Two other typical configurations are given by
\begin{equation}
  \begin{split}
    \phidd (\varphi)
    &= \deltadd
    \begin{pmatrix}
      g(r)  & 0 & 0 \\ 0 &  f(r) e^{i\varphi}  & 0 \\ 0 & 0 & g(r)
    \end{pmatrix}\,,\\
    U(\varphi) &= e^{i(\varphi/6)\diag(-1,2,-1)}\,,\\
    A_i &= - \frac{a(r)}{6 g}\frac{\epsilon_{ij} x^j}{r^2} 
    \diag(-1,2,-1)\,,
  \end{split}
  \label{eq:ansatz2}
\end{equation}
and 
\begin{equation}
  \begin{split}
    \phidd (\varphi)
    &= \deltadd
    \begin{pmatrix}
      f(r) e^{i\varphi}   & 0 & 0 \\ 0 & g(r) & 0 \\ 0 & 0 & g(r)
    \end{pmatrix}\,,\\
    U(\varphi) &= e^{i(\varphi/6)\diag(2,-1,-1)}\,,\\
    A_i &=- \frac{a(r)}{6 g}\frac{\epsilon_{ij} x^j}{r^2} 
    \diag(2,-1,-1)\,.
  \end{split}
  \label{eq:ansatz3}
\end{equation}
We denote configurations in Eqs.~\eqref{eq:ansatz},
\eqref{eq:ansatz2} and \eqref{eq:ansatz3} by 
fluxes that they carry, that is, 
$b$, $g$, and $r$,
respectively.

These three configurations are just typical configurations 
among a more general continuous family of configurations
\begin{align}
  \begin{split}
    \phidd (\varphi)
    &= e^{i\varphi /3} U^O(\varphi) \phidd(0) 
    U^{OT}(\varphi) \\
    &= \deltadd \,O
    \begin{pmatrix}
      g(r)  & 0 & 0 \\ 0 & g(r) & 0 \\ 0 & 0 & f(r) e^{i\varphi}
    \end{pmatrix}
    O^T \,,\\
    U^O(\varphi) &=  O e^{i(\varphi/6)\diag(-1,-1,2)} O^T \\
    &= e^{i(\varphi/6)O \diag(-1,-1,2) O^T}\,, \\
    A_i &= - \frac{a(r)}{6 g}\frac{\epsilon_{ij} x^j}{r^2} 
    O \diag(-1,-1,2) O^T\,, \\
  \end{split}
  \label{eq:ansatz-general}
\end{align}
with $O  \in \mathrm{SO(3)_C}.$ 
This continuous degeneracy can be interpreted as 
zero modes as follows. 
At spatial infinity, the residual symmetry $H = \mathrm{SO(3)} \rtimes
\mathbb{Z}_6$ of the vacuum in the $dd$ phase should be respected.
Around the core of a non-Abelian vortex, however, the symmetry is
further broken spontaneously 
 down to
\begin{align}
  \tilde{K}_{\rm vortex} = \mathrm{O(2)_C} \times (\mathbb{Z}_6)_{\rm
    C+B}\, 
  \label{eq:tilde-K}
\end{align}
by the ansatz in Eq.~\eqref{eq:ansatz} where  
$f \neq g$ around the core.
The symmetry $\mathrm{O(2)_C}$ can be decomposed as
\begin{align}
  \mathrm{O(2)_C} = {\mathbb Z}_2 \ltimes {\rm SO(2)}_{\rm C}\,,
\end{align}
where the SO(2)$_{\rm C}$ is given by
\begin{align}
  \begin{pmatrix}
    \cos \alpha   & -\sin \alpha & 0 \\ \sin \alpha & \cos \alpha & 0 
    \\ 0 & 0 & 1
  \end{pmatrix} 
  \in {\rm SO(2)}_{\rm C},
\end{align}
and ${\mathbb Z}_2$ is generated by $\diag(1,-1,-1)$, $\diag(-1,1,-1)$
or their linear combinations for the configuration in
Eq.~\eqref{eq:ansatz}.
Note that $(\mathbb{Z}_6)_{\rm C+B}$ symmetry remains unbroken because
its action given in Eq.~\eqref{eq:z6} is represented with the block
diagonal matrix of the form:
\begin{align}
  X^k=\begin{pmatrix}
    \omega^k & 0 & \vline & 0 \\
     0 & \omega^k & \vline &  0\\ \hline
     0 & 0 & \vline & \omega^{-2k}
  \end{pmatrix}\,.
\end{align}
Consequently, there appear Nambu-Goldstone
(NG) modes associated with this spontaneous symmetry breaking in the
vicinity of the vortex:
\begin{align}
  O \in \frac{H_0}{\tilde{K}_{\rm vortex}} = \frac{{\rm SO(3)}_{\rm
      C}\rtimes \mathbb{Z}_6}{{\rm O(2)}_{\rm C} \times \mathbb{Z}_6} \simeq
  S^2 / \mathbb{Z}_2 \simeq \mathbb{R} P^2. \label{eq:Alice-moduli}
\end{align}
In the vortex core, there remains the unbroken gauge symmetry \footnote{
This is in contrast to the case of a non-Abelian vortex in the CFL
phase, accompanied by the $\mathbb{C} P^2$ orientational moduli 
\cite{Nakano:2007dr,Eto:2009bh,Eto:2013hoa,Eto:2009tr}.}.

\section{Properties of non-Abelian Alice strings}
\label{sec:properties}

In this section, we study some characteristic properties 
of non-Abelian Alice strings, namely topological obstruction, AB
phases, and decay of a U(1)$_{\rm B}$ vortex into three Alice strings.

\subsection{Topological obstruction of a non-Abelian Alice string}

In this subsection, we will turn to the properties of non-Abelian
Alice string.
In the presence of the string, the unbroken little group $H_\varphi$
is position (in this case the angle $\varphi$) dependent.  These
little groups are related by the gauge transformation $U(\varphi)$
given in Eq.~\eqref{eq:holonomy}
\begin{align}
  H_\varphi = U(\varphi) H_{\varphi = 0}\, U^{-1}(\varphi)
\end{align}
and all of them are isomorphic to the one at $\varphi = 0$,
$H_{\varphi = 0} = H_{dd} = \mathrm{SO(3)} \rtimes \mathbb{Z}_6$.
Therefore, for whole group $H_{\varphi = 2\pi} \simeq H_{\varphi = 0}$.
However, it is not true for an individual generator of $H_\varphi$ 
\cite{Alford:1990mk, Alford:1990ur, Preskill:1990bm}.
The fact that the embedding of $H_\varphi$ in $G_{\rm QCD}$ is
position dependent prevents the global and continuous definition of
the generators of $H_\varphi$ for all values of the angle $\varphi$.

Consider the generators of $\mathrm{SO(3)}$, $T_x$,  $T_y$  and $T_z$.
Let us define 
\begin{align}
  T_{x,y,z}(\varphi) \equiv U(\varphi) T_{x,y,z} U^{-1}(\varphi)\,,
\end{align}
then we can show 
\begin{align}
 T_{x,y}(\varphi = 2\pi) &= - T_{x,y} \neq T_{x,y}(\varphi = 0)\, \nonumber\\
 T_z (\varphi = 2\pi) &= + T_z =  T_z (\varphi = 0)
  \label{eq:encircling-Alice}
\end{align}
for non-Abelian Alice string with the flux of the color $b$ in
Eq.~(\ref{eq:ansatz}).  
We can only recover the original $T_{x,y}$ by
rotating $\varphi = 4\pi$:
\begin{align}
  T_{x,y}(\varphi = 4\pi) 
  = T_{x,y}(\varphi = 0)\,.
  \label{eq:encircling-Alice-twice}
\end{align}

Likewise, similar nonsingle-valuedness are present around Alice strings with
different fluxes:
\begin{align}
  T_{z,x}(\varphi = 2\pi) &= - T_{z,x} \neq T_{z,x}(\varphi = 0)\, \nonumber\\
  T_y (\varphi = 2\pi) &= + T_y =  T_y (\varphi = 0)
  \label{eq:encircling-Alice2}
\end{align}
for non-Abelian Alice string with the flux of the color $g$ in
Eq.~(\ref{eq:ansatz2}), and
\begin{align}
 T_{y,z}(\varphi = 2\pi) &= - T_{y,z} \neq T_{y,z}(\varphi = 0)\, \nonumber\\
 T_x (\varphi = 2\pi) &= + T_x =  T_x (\varphi = 0)
  \label{eq:encircling-Alice3}
\end{align}
for non-Abelian Alice string with the flux of the color $r$ in
Eq.~\eqref{eq:ansatz3}.

This phenomenon that we have shown here is the so-called
\textit{topological obstruction} 
 \cite{Alford:1990mk, Alford:1990ur, Preskill:1990bm}.  
 One of the prominent example of the
topological obstruction can be found in the theory with the string
configuration dubbed an \textit{Alice string}
for which a U(1) generator flips its sign after complete encirclement 
of the string 
\cite{Schwarz:1982ec, Alford:1990mk, Alford:1990ur, Preskill:1990bm,
  Chatterjee:2017jsi, Chatterjee:2017hya,
  Chatterjee:2019zwx}\footnote{In
  Refs.~\cite{Chatterjee:2017jsi,Chatterjee:2017hya,Chatterjee:2019zwx},
  an SU(2) $\times$ U(1) gauge theory with charged triplet scalar
  fields were studied, which is an SU(2) version of our case where
  U(1) is also gauged.  The case of a global U(1) symmetry, closer to
  our case, was discussed in Refs.~\cite{Sato:2018nqy,Chatterjee:2019rch} 
  in the context of the axion dark matter model.}.
We have shown here that the vortex that exists in the $\langle dd
\rangle$ phase in dense QCD is in fact the non-Abelian analog of the
Alice string.  Hereafter, we will call this vortex configuration as a
\textit{non-Abelian Alice string}.

Because of the topological obstruction, one can state that the unbroken
symmetry $H$ of the ground state is further ``broken'' to its subgroup
$\tilde K={\rm O(2)} \times \mathbb{Z}_6$ whose generators are
all single-valued.
Eventually, this breaking is the same with spontaneous
breaking in the vicinity of the vortex in Eq.~\eqref{eq:tilde-K}.

\subsection{Generalized Aharonov-Bohm phases around vortices}
\label{sec:abphase}

\begin{table*}[t]
  \centering
  \begin{tabular}{c|c|c||cc|cc|cc}
    Vortex & \begin{tabular}{c} $\mathrm{U(1)_B}$
      winding\\ $B$ \end{tabular}
    & \begin{tabular}{c} Color-magnetic\\ flux
        $\mathcal{F}$ \end{tabular}
    & \begin{tabular}{c} Generalized \\ AB phase for
        \\ ($u,d$)-quarks \end{tabular} &
    & \begin{tabular}{c} AB phase for\\ an $s$-quark \end{tabular} & &
    \begin{tabular}{c} Generalized\\ AB
      phase for $\phiudop$ \end{tabular}  
    &  \\ \hline \hline
    Abelian vortex & 1 & 0 & $(-1, -1, -1)$ & $\mathbb{Z}_2$ &
    $(1,1,1)$ & $1$ & $(1, 1, 1)$ & $1$\\ \hline
    \begin{tabular}{c}
      Non-Abelian\\
      Alice string
    \end{tabular}
    & 1/3 & 1/6 & 
    $\left(\begin{tabular}{ccc}
      $-1$ & $+1$ & $+1$ \\
      $+1$ & $-1$ & $+1$ \\
      $+1$ & $+1$ & $-1$  \\
    \end{tabular}\right)$ 
    & $\mathbb{Z}_2$ &
    $\left(\begin{tabular}{ccc}
      $\omega^2$    & $\omega^{-1}$ & $\omega^{-1}$   \\
      $\omega^{-1}$ & $\omega^2$    & $\omega^{-1}$ \\
      $\omega^{-1}$ & $\omega^{-1}$ & $\omega^2$  \\
    \end{tabular}\right)$ 
    & $\mathbb{Z}_6$ & 
    $\left(\begin{tabular}{ccc}
      $+1$ & $-1$ & $-1$ \\
      $-1$ & $+1$ & $-1$ \\
      $-1$ & $-1$ & $+1$  \\
    \end{tabular}\right)$ 
    & ${\mathbb{Z}}_2$  
  \end{tabular}
  \caption{AB phases of light $(u,d)$ quarks, heavy ($s$) quark, 
  and the 2SC condensate $\phiud \sim ud$ around various vortices. 
  For row vectors, their columns represent the colors of the quarks or $\phiud$.
  For $3 \times 3$ matrices, rows represent the colors of fluxes of the vortices 
  and columns represent the colors of the quarks or $\phiud$.
  The order $k$ of the $s$-quark AB phase ${\mathbb Z}_k$ corresponds
  to the flux $1/k$ of the vortex.
  \label{tab:abphase0}
  }
\end{table*}

In the case of the CFL phase, the electromagnetic AB phases around a
non-Abelian vortex was calculated~\cite{Chatterjee:2015lbf}. In this
subsection, we consider AB phases of color SU(3)$_{\rm C}$ symmetry
around a vortex, without switching on the electromagnetism.  To this
end, we put the quark field $\qop$, the gauge field $A_i$ and the
2SC diquark operator $\phiudop$ in the above-mentioned vortex
configuration.
When they go around the vortex, they undergo a gauge transformation
according to the Wilson-line integral in Eq.~\eqref{eq:holonomy}.
Thus, after a complete encirclement of the vortex, the fields may pick up a
phase coming from the Wilson loop and the baryon circulation in
$\mathrm{U(1)_B}$.  We call the phase without the baryon circulation
as AB phase and the one with the baryon circulation as
generalized AB phase.
In the later sections, we use this (generalized) AB phase as
a guiding principle to ensure the consistency between the different
fields.

For any $\varphi \neq 0$, the quark field operator $\qop$ and the diquark
operator $\phiudop$ at $\varphi$ are
given by a holonomy action as
\begin{align}
  \qop(\varphi) &\sim e^{i\thetab(\varphi)} U(\varphi) \qop(\varphi = 0)\,, \\
  \phiudop(\varphi) &\sim e^{2i\thetab(\varphi)} U^{-1}(\varphi) \phiudop(\varphi = 0)\,,
\end{align}
where $U(\varphi)$ is defined as in Eq.~\eqref{eq:holonomy}.
One can read out the generalized AB phase (in the exponentiated form)
$\Gamma$ from the fields at $\varphi=2\pi$ after encircling around the
vortex, i.e., $\qop(0) \to \qop(2\pi) = \Gamma \qop(0)$ and $\phiudop(0) \to
\phiudop(2\pi) = \Gamma \phiudop(0)$.
As discussed above, our gauge field $A_i$ is proportional to the
diagonal matrix.  So, taking $A_i \propto \diag(-1, -1, 2)$ for
instance, we can explicitly write down 
\begin{align}
  \begin{split}
  \qop(2\pi) &\sim e^{i\theta_{\rm B}(\varphi)} U(2\pi) \qop(0) \\
  &\sim e^{i\pi B} e^{2i\pi \calF \diag(-1, -1, 2)}
  \begin{pmatrix}\qop_{r}(0) \\ \qop_{g}(0) \\ \qop_{b}(0) \end{pmatrix}\,, \\
  \phiudop(2\pi) &\sim e^{2i\theta_{\rm B}(\varphi)} U^{-1}(2\pi) \phiudop(0) \\
  &\sim e^{2i\pi B} e^{-2i\pi \calF \diag(-1, -1, 2)}
  \begin{pmatrix}\phiudop^r(0) \\ \phiudop^g(0) \\ \phiudop^b(0) \end{pmatrix}
  \end{split}
  \label{eq:abphase}
\end{align}
where $B$ and $\calF$ denote the $\mathrm{U(1)_B}$ circulation and the
color-magnetic flux, respectively.
The values of $B$ and $\calF$ for each vortex are tabulated in
Table~\ref{tab:abphase0}.
When an $s$-quark that does not participate in the condensations 
encircles a vortex, it receives only an AB phase of the
color gauge group.  In contrast, 
the light quarks $u$ and $d$, that participate in the condensations, 
and 
$\phiudop$ receive an additional contribution from the baryon number
symmetry $\rm U(1)_B$ other than the usual AB phase of the color gauge
group, because $\phiudop$ itself contains a vortex winding.  The total
phase is a generalized AB phase.

\subsubsection{Abelian string}

Substituting $B=1$ and $\calF=0$ in Eq.~\eqref{eq:abphase}, 
one gets the following AB phases.
The generalized AB phases $\Gamma$ of light ($u,d$) or heavy ($s$)
quarks encircling around an Abelian U(1)$_{\rm B}$ vortex can be
summarized, by using short-hand notation, as
\begin{align}
  \Gamma_{\alpha\beta}^{u,d} (\varphi) &= 
  \bordermatrix{& r & g & b \cr & e^{+i \varphi/2 } & e^{+i \varphi/2
    } & e^{+i \varphi/2 }}\,\label{eq:abelgammaud}\\
  \Gamma_{\alpha\beta}^{s} (\varphi) &= 
  \begin{pmatrix}
    1  & 1 & 1  \\ \end{pmatrix}\,,
\end{align}  
where the columns ($\beta=r,g,b$) denote the colors of the light
($u,d$) or heavy ($s$) quarks encircling the vortex as explicitly
shown in Eq.~\eqref{eq:abelgammaud}.
After the complete encirclement $\varphi=2\pi$, these phases become
\begin{align}
  \Gamma_{\alpha\beta}^{u,d} (\varphi=2\pi) &= 
  \begin{pmatrix}
    -1 & -1 & -1
  \end{pmatrix}\,,\\
  \Gamma_{\alpha\beta}^{s} (\varphi=2\pi) &= 
  \begin{pmatrix}
    +1    & +1 & +1
  \end{pmatrix}\,.
\end{align}  

On the other hand, when the 2SC condensate $\phiud$ encircles the
vortex, its AB phases are
\begin{align}
  \Gamma_{\alpha\beta}^{\rm 2SC} (\varphi)= 
  \begin{pmatrix}
    e^{+i \varphi } & e^{+i \varphi } & e^{+i \varphi}
  \end{pmatrix}\,.
\end{align}
After the complete encirclement $\varphi=2\pi$, these phases become
\begin{align}
  \Gamma_{\alpha\beta}^{\rm 2SC} (\varphi=2\pi)= 
  \begin{pmatrix}
    +1 & +1 & +1
  \end{pmatrix}\,.
\end{align}

\subsubsection{Non-Abelian Alice string}

Substituting $B=1/3$ and $\calF=1/6$ in Eq.~\eqref{eq:abphase}, one
gets the followings.
The asymptotic gauge fields of a color flux with a color $r,g,b$ are
given by
$A^r_i \propto \diag(2, -1, -1)$, 
$A^g_i \propto \diag(-1, -2, -1)$, 
$A^b_i \propto \diag(-1, -1, 2)$, 
respectively.

Therefore, the pure AB phases of heavy ($s$) quark encircling around
flux tubes can be summarized, again by using short-hand notation, as
\begin{align}
  \Gamma_{\alpha\beta}^{s} (\varphi)=
  \bordermatrix{ & r & g & b \cr
    r & e^{+i \varphi/3 } & e^{-i \varphi/6 } & e^{-i \varphi/6 } \cr
    g & e^{-i \varphi/6 } & e^{+i \varphi/3 } & e^{-i \varphi/6 } \cr
    b & e^{-i \varphi/6 } & e^{-i \varphi/6 } & e^{+i \varphi/3 }} \,,
\end{align}
where, as explicitly indicated above, the row ($\alpha=r,g,b$) denotes
the color of the flux tubes, and the column ($\beta=r,g,b$) denotes the
colors of the heavy ($s$) quark encircling them.
After the complete encirclement $\varphi=2\pi$, these phases become 
\begin{align}
\Gamma_{\alpha\beta}^{s} (\varphi=2\pi)= 
\begin{pmatrix}
          \omega^2    & \omega^{-1} & \omega^{-1} \\
          \omega^{-1} & \omega^2    & \omega^{-1} \\
          \omega^{-1} & \omega^{-1} & \omega^2
       \end{pmatrix}\,,
\end{align}  
which are a color nonsinglet.

When the light quarks $u,d$ encircle the Alice string, they also
receive $\rm U(1)_B$ transformation $e^{+i\varphi/6}$ as well as the
AB phase that they have in common with those of the $s$-quarks.  Therefore,
generalized AB phases of the light quarks $u,d$ are given by
\begin{align}
  \Gamma_{\alpha\beta}^{u,d} (\varphi)
  &= e^{+i\varphi/6} \Gamma_{\alpha\beta}^{s} (\varphi) \notag \\
  &=
  \begin{pmatrix}
    e^{+i \varphi/2 } & 1 & 1   \\
    1 & e^{+i \varphi/2 } & 1 \\
    1 & 1 & e^{+i \varphi/2 }   \\
  \end{pmatrix}\,.
\end{align}  
After the complete encirclement $\varphi=2\pi$, these phases become 
\begin{align}
\Gamma_{\alpha\beta}^{u,d} (\varphi=2\pi)= 
\begin{pmatrix}
          -1 & +1 & +1 \\
          +1 & -1 & +1 \\
          +1 & +1 & -1  \\
       \end{pmatrix}
\end{align}  
which are a color nonsinglet as well.

On the other hand, when the 2SC condensate $\phiud$ encircles the
Alice string, its generalized AB phases are
\begin{align}
  \Gamma_{\alpha\beta}^{\rm 2SC} (\varphi)
  &= e^{+i\varphi/3} \Gamma_{\alpha\beta}^{s} (\varphi) \notag \\
  &=
  \begin{pmatrix}
    1 & e^{+i \varphi/2 } & e^{+i \varphi/2 }   \\
    e^{+i \varphi/2 } & 1 & e^{+i \varphi/2 }   \\
    e^{+i \varphi/2 } & e^{+i \varphi/2 } & 1
  \end{pmatrix}\,.
\end{align}
After the complete encirclement $\varphi=2\pi$, these phases become 
\begin{align}
  \Gamma_{\alpha\beta}^{\rm 2SC} (\varphi=2\pi)= 
  \begin{pmatrix}
    +1 & -1 & -1 \\
    -1 & +1 & -1 \\
    -1 & -1 & +1  \\
  \end{pmatrix}\,,
\end{align}     
which are a color nonsinglet.

The (generalized) AB phase of the $u,d,s$ quarks and the 2SC condensate $\phiud$
are different among the colors, so they are color nonsinglet.
Surprisingly, by winding the quarks or the 2SC condensate around the
string, one can read out the color of the flux 
from infinite distance \footnote{
This is in contrast to the case of non-Abelian vortices (color flux tubes) in 
the CFL phase, around which 
all (generalized) AB phases are color singlet 
\cite{Chatterjee:2018nxe, *Chatterjee:2019tbz}.
  }.

\subsection{Decay of an Abelian U(1)$_{\rm B}$ vortex}
\label{sec:decay}

Here we discuss that a single Abelian U(1)$_{\rm B}$ superfluid vortex
decays into a set of three non-Abelian Alice strings with total color
magnetic flux canceled out, as the case of the CFL
phase~\cite{Nakano:2007dr,Cipriani:2012hr,
Alford:2016dco,
Chatterjee:2018nxe, *Chatterjee:2019tbz}.

\begin{figure}
  \centering
  \begin{tabular}{cc}
    \includegraphics[width=.45\linewidth,keepaspectratio]{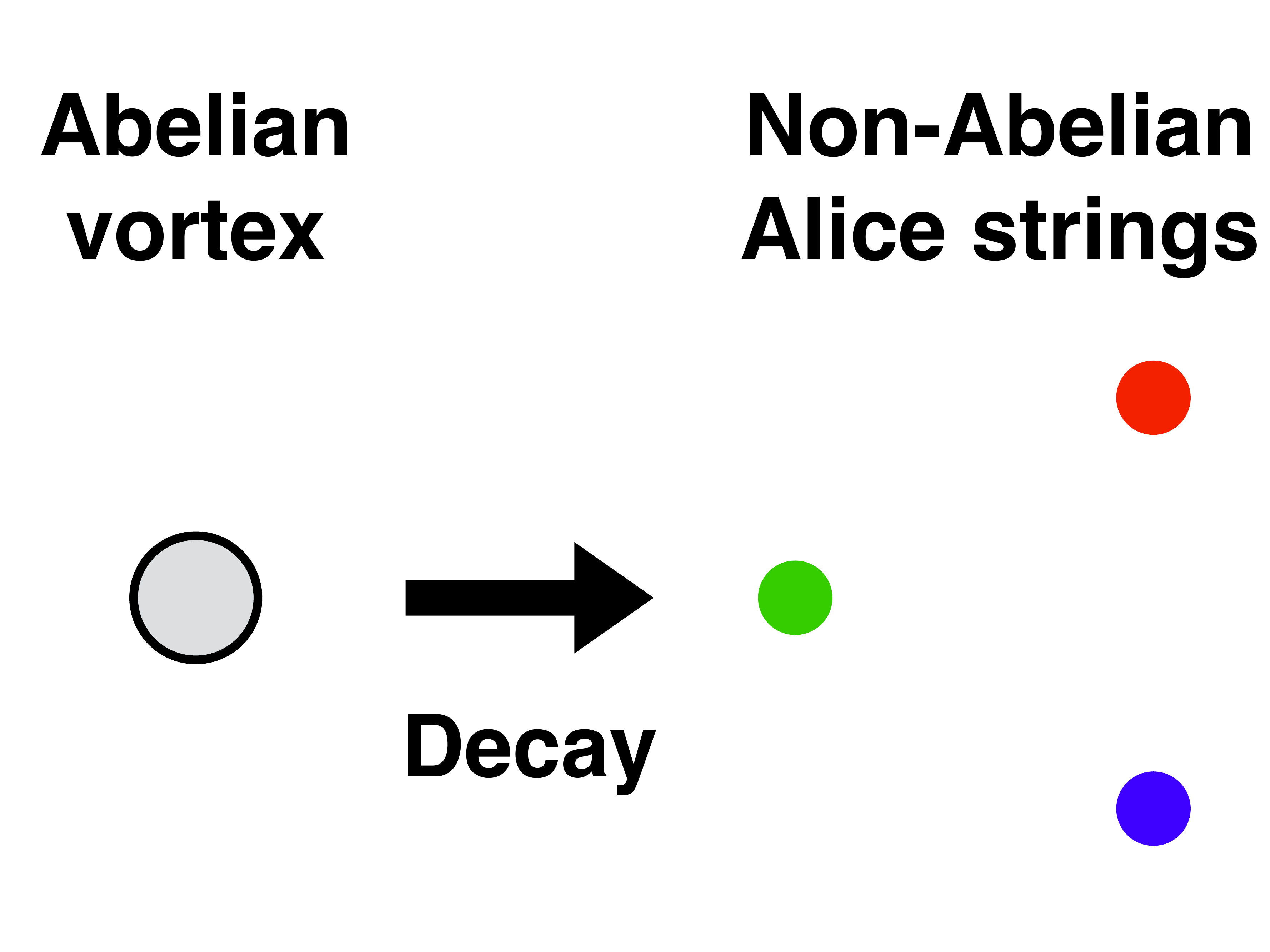} &
    \includegraphics[width=.55\linewidth,keepaspectratio]{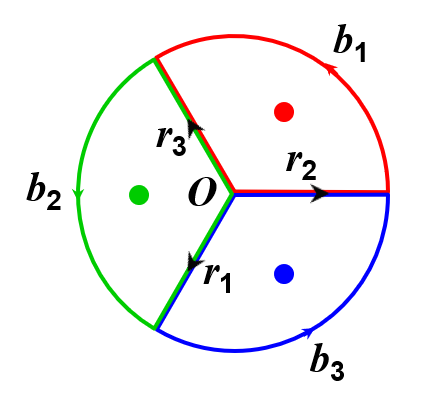}
    \\
    (a) & (b)
  \end{tabular}
  \caption{\label{fig:vortex-decay1} 
    Detailed configurations of decays of a U(1)$_{\rm B}$ superfluid
    vortex.  (a) Schematic illustration of the decay of an Abelian vortex 
    into three non-Abelian Alice strings.
    (b) The configurations after decay.  The U(1)$_{\rm B}$ vortex
    initially located at the origin $O$ decays into three non-Abelian
    Alice strings, denoted by the red, green, blue blobs.  The $b_1$,
    $b_2$ and $b_3$ are the paths with angles $2\pi/3$ at the boundary
    at the spatial infinity, and $r_1$, $r_2$, $r_3$ denote the paths
    from the origin $O$ to spatial infinities.
}
\end{figure}

Superfluid vortices have logarithmically divergent tension (energy per
unit length), $E \sim n^2 \log \Lambda$, in a system of the size
$\Lambda$.  Here, $n$ is the winding number of U(1)$_{\rm B}$, or the
superfluid circulation.  An Abelian U(1)$_{\rm B}$ vortex in
Eq.~(\ref{eq:Abelian}) has $n=1$ while a non-Abelian Alice string in
Eq.~\eqref{eq:ansatz}, \eqref{eq:ansatz2}, \eqref{eq:ansatz3} or more
generally Eq.~\eqref{eq:ansatz-general} has $n=1/3$.
Thus, this fact implies a decay of Abelian vortices into a set of
three non-Abelian vortices as schematically shown in
Fig.~\ref{fig:vortex-decay1}(a), because of energetics
\begin{align}
 E(n=1) = 9 E(n=1/3) > 3 E(n=1/3)\,.
\end{align}
If three non-Abelian vortices are infinitely separated, the total
energy is proportional to $3 E(n=1/3)$, which is 1/3 of the energy of a
single Abelian vortex.

The U(1)$_{\rm B}$ vortex has no color-magnetic flux while non-Abelian
Alice strings in Eq.~\eqref{eq:ansatz}, \eqref{eq:ansatz2},
\eqref{eq:ansatz3} do with their sum canceled out.
We thus conclude
\begin{align}
  \phidd = \phidd^{\rm U(1)_B} & \;\; \to \;\;
  \phidd = \phidd^r  +  \phidd^g  +  \phidd^b\,, \\
  A_i = A_i^{\rm U(1)_B} = 0 & \;\; \to \;\; A_i = A_i^r + A_i^g + A_j^b\,, 
\end{align}
with
\begin{align}
  \begin{split}
    \phidd^{\rm U(1)_{\rm B}} 
    &= \deltadd\, \diag(f_0(r) e^{i\varphi}, f_0(r) e^{i\varphi}, f_0(r)
    e^{i\varphi})\,,\\
    \phidd^r 
    &= \deltadd\, \diag(g(r), g(r), f(r) e^{i\varphi})\,,\\
    \phidd^g 
    &= \deltadd\, \diag(g(r), f(r) e^{i\varphi}, g(r))\,,\\
    \phidd^b 
    &= \deltadd\, \diag(f(r) e^{i\varphi}, g(r), g(r))\,,\\
    A_i^r &= - \frac{a^r(r)}{6 g}\frac{\epsilon_{ij} x_1^j}{r_1^2} 
    \diag(-1,-1,2)\,, \\
    A_i^g &= - \frac{a^g(r)}{6 g}\frac{\epsilon_{ij} x_2^j}{r_2^2} 
    \diag(-1,2,-1)\,, \\
    A_i^b &= - \frac{a^b(r)}{6 g}\frac{\epsilon_{ij} x_3^j}{r_3^2} 
    \diag(2,-1,-1)\,.
  \end{split}
\end{align}

Let us check the consistency in terms of the generalized AB phases 
and topological obstructions.
To this end,
we divide a closed loop at infinity encircling the U(1)$_{\rm B}$
vortex to $b_1$, $b_2$, and $b_3$ as in Fig.~\ref{fig:vortex-decay1}(b).
Along the paths $b_1$, $b_2$, and $b_3$, the U(1)$_{\rm B}$ symmetry
acts as $\exp(i \varphi/3)$ giving rise to a $\omega^2$ factor so that
each path carries 1/3 circulation and the whole loop $b_1+b_2+b_3$
carries the unit circulation of the U(1)$_{\rm B}$ vortex.
We define radial paths $r_1$, $r_2$ and $r_3$ starting at the origin
to infinities as in Fig.~\ref{fig:vortex-decay1}.
The closed loop $l_r = b_1 -r_3  + r_2$, $l_g =b_2 -r_1 + r_3$, and
$l_b = b_3 - r_2 + r_1$ enclose $r$, $g$ and $b$ non-Abelian Alice
strings, respectively.

Then we find that the SU(3)$_{\rm C}$ gauge transformations
along the radial paths $r_1$, $r_2$, and $r_3$ are given by 
\begin{align}
  \begin{split}
    r_1:& \quad U = \exp (iF(r)/6) \diag(1,-1,0)\,, \\
    r_2:& \quad U = \exp (iF(r)/6) \diag(-1,0,1)\,, \\
    r_3:& \quad U = \exp (iF(r)/6) \diag(0,1,-1)\,,
  \end{split}
  \label{eq:r-action}
\end{align}
respectively, with an arbitrary function $F(r)$ satisfying the
boundary conditions
\begin{align}
  F(r=0)=0, \quad F(r=\infty) = 2\pi\,.
\end{align}
Then, the loops $l_r$, $l_g$, and $l_b$ enclosing $r$, $g$, and $b$
non-Abelian Alice strings correctly recover the original group actions
of the following elements $(e^{i \varphi/3}, U) \in {\rm U(1)_B \times
  SU(3)_C}$:
\begin{align}
  \begin{split}
    l_r = b_1 -r_3  + r_2: & \quad (e^{i \varphi/3}, U=e^{iF/6}\diag(-1,-1,2))\,, \\
    l_g = b_2 -r_1 + r_3: & \quad  (e^{i \varphi/3}, U=e^{iF/6}\diag(-1,2,-1))\,, \\
    l_b = b_3 - r_2 + r_1: & \quad (e^{i \varphi/3}, U=e^{iF/6}\diag(2,-1,-1))\,,
  \end{split}
\end{align}
respectively.

When we encircle three non-Abelian strings ($r,g,b$), total AB phases
of any particles coincide with those of a $\rm U(1)_B$ vortex, as
can be seen from Table~\ref{tab:abphase0}.
For $u, d$ quarks, if one multiply all three elements of each column
(color of $u,d$ quarks) of the third line of Table~\ref{tab:abphase0},
one get $(+1)(+1)(-1)=-1$ for all colors of $u,d$ quarks, which
coincides with the first line for a $\rm U(1)_B$ vortex.
The same holds for a $s$ quark $\omega^2 \omega^{-1} \omega^{-1} = 1$
for all colors of the $s$ quark.

Furthermore, although each non-Abelian string exhibits topological
obstruction given in Eq.~\eqref{eq:encircling-Alice},
\eqref{eq:encircling-Alice2}, or \eqref{eq:encircling-Alice3}, the
total topological obstruction is canceled out when we encircle all
the three non-Abelian strings ($r,g,b$).
Thus, a topological obstruction is dynamically created under the decay
of a $\rm U(1)_B$ string.

\section{Turning on the 2SC condensate, and bulk-soliton moduli locking}
\label{sec:bulksoliton}

So far we have seen the vortex configurations in the presence of the
$\langle dd \rangle$ condensate $\phidd$.  In particular, there
appear the non-Abelian vortices, which can be regarded as a
non-Abelian extension of the so-called Alice string.
Here, in this section, we will turn on the 2SC condensate $\phiud$ in
this vortex background, and show that the 2SC condensate $\phiud$ will
be aligned toward the direction of the flux.

\subsection{Single non-Abelian Alice string}
As mentioned in Sec.~\ref{sec:SSB2SC}, one can use the remaining gauge
degrees of freedom in the presence of $\phidd$ to transform $\phiud$
into one component.
When turning on $\phiud$, therefore we can take ansatze of the form
$\phiud = (\deltaud, 0, 0)^T$, $(0, \deltaud, 0)^T$ or $(0, 0,
\deltaud)^T$.  We will have to consider their gauge rotation
consistently with the $\dd$ condensate $\phidd$.
In the presence of a single non-Abelian Alice string of
Eq.~\eqref{eq:ansatz}, $\phiud$ receives generalized AB phases as (see
Sec.~\ref{sec:abphase})
\begin{align}
  \phiud &\to e^{i \varphi/3} U^{-1}(\varphi) \phiud \nonumber \\
  & =
  \begin{pmatrix}
    e^{i\varphi/2} & 0 & 0 \\
    0 & e^{i\varphi/2} & 0 \\
    0 & 0 & 1
  \end{pmatrix} \phiud \notag\\
  &=
  \begin{cases}
    (e^{i\varphi/2} \deltaud, 0, 0)^T \\
    (0, e^{i\varphi/2} \deltaud, 0)^T \\
    (0, 0, \deltaud)^T
  \end{cases}
\end{align}
for $\phiud = (\deltaud, 0, 0)^T$, $(0, \deltaud, 0)^T$ or $(0, 0,
\deltaud)^T$, respectively.  
This explicitly shows that at $\varphi = 2\pi$, if the first or the
second component of $\phiud$ is nonzero, they would not be
single-valued, to be inconsistent.  Therefore, it cannot be realized.  One is only left
with the possibility where the third component of $\phiud$ is nonzero,
i.e., $\phiud = (0, 0, \deltaud)^T$, and it implies that in the
presence of the non-Abelian vortex of $dd$ diquark condensate, the 2SC
condensate must be aligned along the direction of the non-Abelian
moduli from the consistency.

This alignment of two condensates should be compared with the zero
mode appearing in the vicinity of the vortex, the Alice string moduli
${\mathbb R}P^2$ in Eq.~\eqref{eq:Alice-moduli}.
Apparently,  in the presence of an Alice string in
Eq.~\eqref{eq:ansatz2} or \eqref{eq:ansatz3}, $\phiud$ must be 
$\phiud = (0,\deltaud, 0)^T$ or $\phiud = (\deltaud, 0, 0)^T$,
respectively.  Thus, the expectation value of the 2SC condensate
$\phiud$ is aligned to the direction of color-magnetic flux or the
moduli of the Alice string.
We call this phenomenon as a ``bulk-soliton moduli locking''.
Conversely, since the 2SC condensate $\phiud$ is the bulk quantity, it
may be natural to say that rather the Alice string moduli ${\mathbb
  R}P^2$ are locked to $\phiud$.   In this picture, the Alice string is
``Abelianized,'' and if there are several Alice strings, all Alice
string moduli are aligned to $\phiud$.

\subsection{Decay of an Abelian $\mathrm{U(1)_B}$ vortex}

\begin{figure*}
  \centering
  \begin{tabular}{cc}
    \includegraphics[width=0.5\linewidth,keepaspectratio]{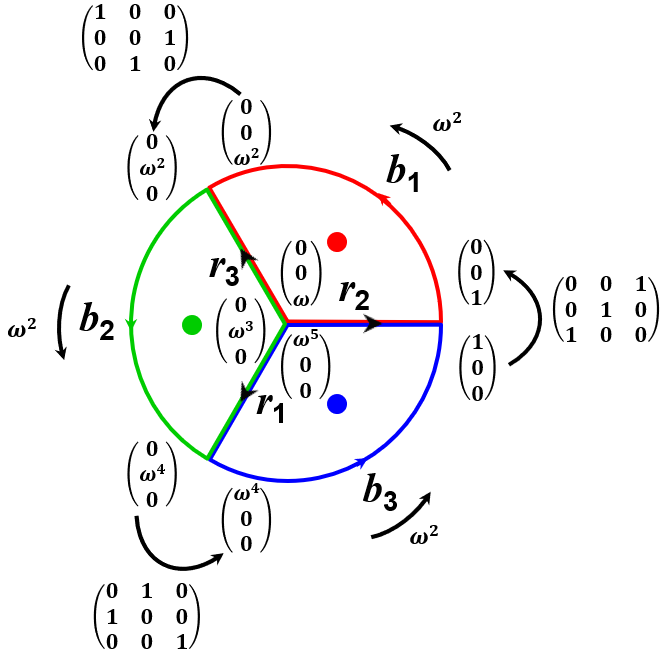}&
    \includegraphics[width=0.4\linewidth,keepaspectratio]{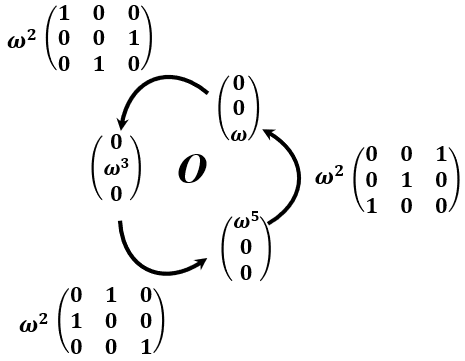}
    \\
    (a) & (b)
  \end{tabular}
  \caption{\label{fig:vortex-decay2} 
    Detailed configurations of decays of a $U(1)_{\rm B}$ superfluid
    vortex in the presence of the 2SC condensate $\phiud$.  All row
    vectors denote $\phiud$.
    (a) The $U(1)_{\rm B}$  vortex initially located at the origin $O$
    decays into three non-Abelian Alice strings, denoted by the red,
    green, blue blobs.  The $b_1$, $b_2$ and $b_3$ are the paths with
    angles $2\pi/3$ at the boundary at the spatial infinity, and
    $r_1$, $r_2$, $r_3$ denote the paths from the origin $O$ to
    spatial infinities.
    (b) Closer look around the origin.
}
\end{figure*}
One may wonder whether the 2SC condensate $\phiud$ might prevent the
decay of a U(1)$_{\rm B}$ superfluid vortex into three non-Abelian
Alice strings (see discussion in Sec.~\ref{sec:decay}), since it is a
nontrivial question whether the 2SC condensate $\phiud$ can be locked
to each non-Abelian vortex consistently to cover the entire space.
In this subsection, we show that the decay of a U(1)$_{\rm B}$ vortex
is still possible in the presence of $\phiud$.

Let us discuss the configuration of $\phiud$ in the configuration of
the U(1)$_{\rm B}$ vortex decay in Fig.~\ref{fig:vortex-decay1}.
To this end, it is convenient to take a gauge of $\phiud$ for the
initial $\rm U(1)_B$ vortex as
\begin{align}
  \begin{split}
    \phidd &\sim e^{i \varphi} \deltadd {\bf 1}_3\,,\\
    \phiud &\sim \frac{\deltaud}{\sqrt3} e^{i\varphi} (1\ 1\ 1)^T\,.
  \end{split}
\end{align}
Then, the configuration of $\phiud$ after decay is almost fixed by the
gauge transformation in Eq.~\eqref{eq:r-action} along the  radial
paths $r_1$, $r_2$, and $r_3$, and $\rm U(1)_B$ transformation
$\omega^2$ along the boundary paths $b_1$, $b_2$, and $b_3$ at
infinity.  Another condition is the absence of AB defects around the
non-Abelian Alice strings, fixing nonzero components of $\phiud$
around each non-Abelian Alice string due to the bulk-soliton moduli
locking.
We thus obtain a possible configuration of $\phiud$ as summarized in
Fig.~\ref{fig:vortex-decay2}(a).
Note that while the phase of $\phiud$ varies along the radial paths
$r_1$, $r_2$, and $r_3$, and the boundary paths $b_1$, $b_2$, and $b_3$
at infinity, the overall phase rotation along a closed loop encircling
each non-Abelian string is canceled out.
In order to glue the boundaries between two neighboring Alice strings,
one needs gauge transformations to change the position of the nonzero
component of $\phiud$.
The remaining nontrivial point appears around the origin as zoomed up
in Fig.~\ref{fig:vortex-decay2}(b).
This can be smoothly connected to $\phiud^\alpha =
\frac{\deltaud}{\sqrt3}  (1\ 1\ 1)^T$ at the origin because of the
triviality of $\pi_1$ for the symmetry breaking by $\phiud$:
$\pi_1(G_{\rm QCD}/\tilde{H}_{\rm 2SC})=0$.

We thus conclude that the 2SC condensate $\phiud$ does not prevent the
decay of a U(1)$_{\rm B}$ superfluid vortex into three non-Abelian
Alice strings.

\section{Consistency with the opposite ordering in the symmetry breaking}
\label{sec:consistency}

Now we address vortices appearing in the opposite ordering of the
condensate formations given in Eq.~\eqref{eq:ghtildek}.  Namely, we
now consider the case where the 2SC condensate $\phiud$ forms first, 
followed by the $\dd$ condensate $\phidd$ formation.
The vortices appear corresponding to the $\phidd$ formation because of
the relation $\pi_1(\tilde{H}_{\rm 2SC}/K_{\mathrm{2SC}+dd}) =
\mathbb{Z}$ as already exemplified in Sec.~\ref{sec:opposite}.
The 2SC condensate in the vacuum reads $(\phiud)^\alpha = \deltaud
\delta^{\alpha3}$.

\subsection{Superfluid vortex}

The simplest vortex is a superfluid vortex of the form
\begin{align}
  \begin{split}
    \phidd(\varphi) &= f_0(r) e^{i \varphi} \deltadd {\bf 1}_3\,,\\
    \phiud(\varphi) &= h_0(r) e^{i\varphi} (0\ 0\ \deltaud)^T
  \end{split}
\end{align}
with the boundary conditions
\begin{align}
f_0 (0) = h_0 (0) =0, \quad f_0(\infty) = h_0(\infty)=1\,.
\end{align}
This arises corresponding to the breaking of $\rm U(1)_B$ symmetry,
and is exactly a $\mathrm{U(1)_B}$ vortex considered in
Sec.~\ref{sec:abelian}.

\subsection{Semi-superfluid $U(1)_{\rm C + B}$ vortices}

Here, we consider the vortex with orientational moduli
in the internal non-Abelian gauge space.  The ansatz we impose here is
of several kinds and we show the asymptotic form.

In the presence of the 2SC condensate, the residual symmetry is
$\tilde H_{\rm 2SC} = \mathrm{SU(2)_C \times U(1)_{C+B}}$ as already
given in Eq.~\eqref{eq:htilde}.
The $\rm U(1)_{C+B}$ symmetry in $\tilde H_{\rm 2SC}$: $\{(e^{i\alpha
  T_8}, e^{2i\alpha}) \in \mathrm{SU(3)_C \times U(1)_B}:
T_8=\diag(1,1,-2)\}$ acts on $\phidd$ as
\begin{align}
  \begin{split}
    \phidd &=
    \begin{pmatrix}
      \deltadd & 0 & (\phidd)_{13} \\
      0 & \deltadd' & (\phidd)_{23} \\
      (\phidd)_{31} & (\phidd)_{32} & \deltadd''
    \end{pmatrix} \\
    \to \ &e^{2i\alpha} e^{i\alpha T_8}\, \phidd \, (e^{i\alpha T_8})^T \\
    & = \begin{pmatrix}
      \deltadd & 0 & e^{3i\alpha} (\phidd)_{13} \\
      0 & \deltadd' & e^{3i\alpha} (\phidd)_{23} \\
      e^{3i\alpha} (\phidd)_{31} & e^{3i\alpha} (\phidd)_{32} &
      e^{6i\alpha} \deltadd''
    \end{pmatrix}\,,
  \end{split}
  \label{eq:U(1)C+B-on-M}
\end{align}
while the 2SC condensate $\phiud \to \phiud$ is kept the same.
Whether the off-diagonal blocks $(\phidd)_{13},(\phidd)_{23},
(\phidd)_{31}, (\phidd)_{32}$ exist or not is crucial for
constructions of vortices.  We now restrict ourselves to the case
in which the off-diagonal blocks vanish as in Sec.~\ref{sec:opposite}.
See the forthcoming paper for the full discussion.

When the off-diagonal blocks are absent, winding only resides in the
$(3,3)$ component of $\phidd$.
Thus, we take the dependence of the condensates on the angular
coordinate $\varphi$ as $\varphi = 6\alpha$ in
Eq.~\eqref{eq:U(1)C+B-on-M}:
\begin{align}
  \begin{split}
    \phiud^\alpha &= (0\ 0\ \deltaud)^T\,,\\
    \phidd (\varphi)
    &= e^{i\varphi/3} U(\varphi) \phidd(0) U^T(\varphi)\\
    &= \deltadd
    \begin{pmatrix}
      g(r) & 0    & 0 \\ 
      0    & g(r) & 0 \\ 
      0    & 0    & f(r) e^{i\varphi}
    \end{pmatrix}\,,\\
    U(\varphi) &= e^{i(\varphi/6)\diag(-1,-1,2)}\\
    A_i &= - \frac{a(r)}{6 g}\frac{\epsilon_{ij} x^j}{r^2} 
    \diag(-1,-1,2)
  \end{split}
  \label{eq:ansatz-left2}
\end{align}
with the boundary conditions
\begin{align}
 f (0) = g' (0) =0, \quad f (\infty) =g(\infty) = 1, 
\end{align}
where we have set the condensate at $\varphi = 0$ as 
\begin{align}
  \phidd (\varphi = 0) 
  = \phidd \begin{pmatrix}
    g(r)  & 0 & 0 \\ 0 & g(r) & 0 \\ 0 & 0 & f(r) 
  \end{pmatrix}
  \sim \phidd \boldsymbol{1}_3\,.
\end{align}
This carries $1/6$ quantized color-magnetic flux $\calF_0$ and $1/3$
quantized circulation in $\rm U(1)_B$.
This is precisely a non-Abelian Alice string locked with $\phiud$.

For each of the above-mentioned vortices other typical configurations
are given by the ones in Eqs.~(\ref{eq:ansatz}), (\ref{eq:ansatz2}),
and (\ref{eq:ansatz3}) by $r$, $g$, and $b$, respectively.  Likewise,
these three configurations can be rotated in the color space.

\section{Summary and discussions}
\label{sec:summary}

In this work we have proposed the novel topological solitons,
\textit{non-Abelian Alice strings}, that emerge in two-flavor dense
QCD.
First, we have given a brief review of the new color-superconducting
phase of two-flavor dense quark matter, 2SC+$\dd$ phase, based on the
quark-hadron continuity scenario in connection with the current
neutron star phenomenology~\cite{Fujimoto:2019sxg, *Fujimoto:2020cho}.
There appears a color anti-symmetric diquark condensate, which is the
so-called 2SC condensate [See Eq.~\eqref{eq:2sc}], and a color
symmetric diquark condensate, which is the $\dd$ diquark condensate
[See Eq.~\eqref{eq:dd}].
Then, we have examined the residual symmetries in the presence of
these diquark condensates.  The result is concisely summarized in
Fig.~\ref{fig:SSB}.  We have turned on these diquark condensates in
sequence, firstly the $\dd$ condensate, and then the 2SC condensate,
and have identified the unbroken subgroups [See the right side of
Fig.~\ref{fig:SSB}] as well as the corresponding order parameter
manifolds.  The new feature here is that the unbroken subgroup
$H_{dd}$ has the semidirect product, which leads to the Alice
phenomenon.  We have also done the consistency check with the opposite
ordering of turning on the diquark condensates [See the left side of
Fig.~\ref{fig:SSB}].

Based on these symmetry breaking patterns, we have introduced the
vortices appearing in the presence of $\dd$ diquark.  The central
object in this paper is the non-Abelian Alice string introduced in
Sec.~\ref{sec:alice}.  This carries the fractional winding in $\rm
U(1)_B$ and also a color-magnetic flux.  We have set forth the
characteristic property of this string, i.e., topological obstruction.
We have introduced the notion of the generalized AB phase,
which is the combination of the $\rm U(1)_B$ circulation and the gauge
rotation around the vortex.  Along these lines, we have discussed the
decay of an Abelian $\rm U(1)_B$ vortex.  The Abelian $\rm U(1)_B$
vortex is the simplest object in the superfluid matter, however, it is
not the most stable one.  We have shown on the basis of the energetics
and the generalized AB phase that a single Abelian vortex should be
separated into a triple of the non-Abelian Alice strings.

We have only considered the existence of $\dd$ up to here, but when
turning on the 2SC diquark, then one must take care of the consistency
in terms of the gauge transformation.  Again the guiding principle is
the generalized AB phase.  We have explicitly shown that in the
presence of $\dd$ the form of the 2SC ansatz is restricted and should
be aligned toward the winding component of the $\dd$ condensate.  Then
we have been lead to the observation that the soliton moduli, which are
the degrees of freedom corresponding to the winding component of the
vortex [See also Eq.~\eqref{eq:Alice-moduli}], are actually
\textit{locked} with the bulk quantity, which is nothing but the 2SC
diquark here.  We named this phenomenon as a ``bulk-soliton moduli
locking''.
We have also checked that in the presence of the 2SC condensate, the
decay of a $\rm U(1)_B$ vortex is still possible.

Finally, we have assessed the case when we turned around the ordering
of the diquark formations in order to maintain the consistency.  The
vortices in this case is the same as the ones discussed above.  This
is as we have expected because we will finally arrive at the identical
coexistence phase of the 2SC and $\dd$, so the physics should not
depend on this ordering.

In this paper, we have restricted ourselves to the most symmetric form
of the ansatze for the condensates.  In particular, we have considered
the case of $\Phi_{\rm 2SC} \in \mathbb{R}^3$ in this paper;  The
other possibility $\phiud \in \mathbb{C}^3$ will be discussed
elsewhere.  We have also assumed that eigenvalues of $\dd$ are
degenerate.  Basically, they do not have to be degenerate,
and we can consider more general setup.
Another simplification was the absence of the electromagnetism.  For
the CFL phase, effects of turning on the electromagnetism on vortices
can be incorporated as the effective potential on the ${\mathbb C}P^2$
moduli space~\cite{Vinci:2012mc}.  Such an effect on the ${\mathbb
  R}P^2$ moduli space could be studied for the case of the 2SC+$\dd$
phase as well.

The major motivation to consider the 2SC+$\dd$ phase is the
quark-hadron continuity between two-flavor quark and hadronic matter.
Thus, it is a natural question how vortices in the 2SC+$\dd$ phase
discussed in this paper are connected to those in the $^3P_2$ hadronic
phase.  In the $^3P_2$ phase, there are half-quantized
vortices~\cite{Masuda:2016vak} in addition to integer
vortices~\cite{Muzikar:1980as, Sauls:1982ie, Masuda:2015jka,
  Chatterjee:2016gpm, Masaki:2019rsz}.
It was also shown in Ref.~\cite{Mizushima:2016fbn} that $^3P_2$
superfluids are topological superfluids exhibiting topologically
protected gapless Majorana fermions on their
boundary~\cite{Mizushima:2016fbn} or inside vortex
cores~\cite{Masaki:2019rsz}.
Thus, it is an interesting question whether such gapless fermion modes
exist in vortex cores in 2SC+$\dd$ phase to study validity of
quark-hadron continuity.
As this problem concerns, gapless Majorana fermion modes indeed exist
in non-Abelian vortices in the CFL
phase~\cite{Yasui:2010yw,Fujiwara:2011za}.
Also, in the higher density region, the 2SC+$\dd$ phase may be
connected to the CFL phase.  Thus, it is also an intriguing question
how vortices in the 2SC+$\dd$ phase are connected to those in the CFL
phase.  The strange quark mass will take a crucial part in studying
along this line, and the effect of quark masses has already been
discussed for non-Abelian vortices in the CFL phase~\cite{Eto:2009tr}.
These and related topics are to be explored in future studies.

It is often the case that when quasiparticles exhibit nontrivial AB
phases around a vortex, the system is topologically ordered.  It was,
however, discussed in Refs.~\cite{Hirono:2018fjr, Hirono:2019oup} that
the CFL phase is {\it not} topologically ordered albeit a non-Abelian
nature of the AB phases of vortices.
This is because of superfluidity of the CFL phase.
The same would hold for 2SC+$\dd$ phase discussed in this paper.
Concerning this issue, when U(1)$_{\rm B}$ is coupled to a U(1) gauge
field, the system is topologically ordered~\cite{Hidaka:2019jtv}.
Thus, the 2SC+$\dd$ phase would give a novel topologically ordered
phase if the U(1)$_{\rm B}$ is gauged.

We have studied nontrivial AB phases of quasiparticles around strings. 
In the case of non-Abelian strings, exchanging strings themselves may
give rise to a nontrivial braiding statistics, as discussed in
Ref.~\cite{Lo:1993hp}.
Regarding this, when Majorana fermion zero modes exist in the case of
topological superconductors such as chiral $P$-wave superconductors,
it also gives rise to another kind of non-Abelian
braidings~\cite{Ivanov:2000mjr}.  It is also actually the case for
non-Abelian vortices in the CFL phase~\cite{Yasui:2010yh,
  Hirono:2012ad} due to Majorana fermion zero
modes~\cite{Yasui:2010yw, Fujiwara:2011za}.  These two facts suggest,
in the case of the 2SC+$\dd$ phase, a very interesting possibility of
novel non-Abelian statistics having two different origins.

\begin{acknowledgments}

We thank Shigehiro Yasui for a discussion at the early stage of this work.
Y.~F. thanks Kenji~Fukushima for reading the manuscript and comments.
This work is supported in part by Grant-in-Aid for Scientific
Research, JSPS KAKENHI Grants
No.~JP20J10506 (Y.F.) 
and 
No.~JP18H01217 (M.N.).

\end{acknowledgments}

\bibliographystyle{apsrev4-1}
\bibliography{bib_alice}

\begin{thebibliography}{89}%
\makeatletter
\providecommand \@ifxundefined [1]{%
 \@ifx{#1\undefined}
}%
\providecommand \@ifnum [1]{%
 \ifnum #1\expandafter \@firstoftwo
 \else \expandafter \@secondoftwo
 \fi
}%
\providecommand \@ifx [1]{%
 \ifx #1\expandafter \@firstoftwo
 \else \expandafter \@secondoftwo
 \fi
}%
\providecommand \natexlab [1]{#1}%
\providecommand \enquote  [1]{``#1''}%
\providecommand \bibnamefont  [1]{#1}%
\providecommand \bibfnamefont [1]{#1}%
\providecommand \citenamefont [1]{#1}%
\providecommand \href@noop [0]{\@secondoftwo}%
\providecommand \href [0]{\begingroup \@sanitize@url \@href}%
\providecommand \@href[1]{\@@startlink{#1}\@@href}%
\providecommand \@@href[1]{\endgroup#1\@@endlink}%
\providecommand \@sanitize@url [0]{\catcode `\\12\catcode `\$12\catcode
  `\&12\catcode `\#12\catcode `\^12\catcode `\_12\catcode `\%12\relax}%
\providecommand \@@startlink[1]{}%
\providecommand \@@endlink[0]{}%
\providecommand \url  [0]{\begingroup\@sanitize@url \@url }%
\providecommand \@url [1]{\endgroup\@href {#1}{\urlprefix }}%
\providecommand \urlprefix  [0]{URL }%
\providecommand \Eprint [0]{\href }%
\providecommand \doibase [0]{http://dx.doi.org/}%
\providecommand \selectlanguage [0]{\@gobble}%
\providecommand \bibinfo  [0]{\@secondoftwo}%
\providecommand \bibfield  [0]{\@secondoftwo}%
\providecommand \translation [1]{[#1]}%
\providecommand \BibitemOpen [0]{}%
\providecommand \bibitemStop [0]{}%
\providecommand \bibitemNoStop [0]{.\EOS\space}%
\providecommand \EOS [0]{\spacefactor3000\relax}%
\providecommand \BibitemShut  [1]{\csname bibitem#1\endcsname}%
\let\auto@bib@innerbib\@empty
\bibitem [{\citenamefont {Alford}\ \emph {et~al.}(2008)\citenamefont {Alford},
  \citenamefont {Schmitt}, \citenamefont {Rajagopal},\ and\ \citenamefont
  {Sch{\"a}fer}}]{Alford:2007xm}%
  \BibitemOpen
  \bibfield  {author} {\bibinfo {author} {\bibfnamefont {M.~G.}\ \bibnamefont
  {Alford}}, \bibinfo {author} {\bibfnamefont {A.}~\bibnamefont {Schmitt}},
  \bibinfo {author} {\bibfnamefont {K.}~\bibnamefont {Rajagopal}}, \ and\
  \bibinfo {author} {\bibfnamefont {T.}~\bibnamefont {Sch{\"a}fer}},\ }\href
  {\doibase 10.1103/RevModPhys.80.1455} {\bibfield  {journal} {\bibinfo
  {journal} {Rev. Mod. Phys.}\ }\textbf {\bibinfo {volume} {80}},\ \bibinfo
  {pages} {1455} (\bibinfo {year} {2008})},\ \Eprint
  {http://arxiv.org/abs/0709.4635} {arXiv:0709.4635 [hep-ph]} \BibitemShut
  {NoStop}%
\bibitem [{\citenamefont {Alford}\ \emph
  {et~al.}(1999{\natexlab{a}})\citenamefont {Alford}, \citenamefont
  {Rajagopal},\ and\ \citenamefont {Wilczek}}]{Alford:1998mk}%
  \BibitemOpen
  \bibfield  {author} {\bibinfo {author} {\bibfnamefont {M.~G.}\ \bibnamefont
  {Alford}}, \bibinfo {author} {\bibfnamefont {K.}~\bibnamefont {Rajagopal}}, \
  and\ \bibinfo {author} {\bibfnamefont {F.}~\bibnamefont {Wilczek}},\ }\href
  {\doibase 10.1016/S0550-3213(98)00668-3} {\bibfield  {journal} {\bibinfo
  {journal} {Nucl. Phys. B}\ }\textbf {\bibinfo {volume} {537}},\ \bibinfo
  {pages} {443} (\bibinfo {year} {1999}{\natexlab{a}})},\ \Eprint
  {http://arxiv.org/abs/hep-ph/9804403} {arXiv:hep-ph/9804403} \BibitemShut
  {NoStop}%
\bibitem [{\citenamefont {Alford}\ \emph {et~al.}(1998)\citenamefont {Alford},
  \citenamefont {Rajagopal},\ and\ \citenamefont {Wilczek}}]{Alford:1997zt}%
  \BibitemOpen
  \bibfield  {author} {\bibinfo {author} {\bibfnamefont {M.~G.}\ \bibnamefont
  {Alford}}, \bibinfo {author} {\bibfnamefont {K.}~\bibnamefont {Rajagopal}}, \
  and\ \bibinfo {author} {\bibfnamefont {F.}~\bibnamefont {Wilczek}},\ }\href
  {\doibase 10.1016/S0370-2693(98)00051-3} {\bibfield  {journal} {\bibinfo
  {journal} {Phys. Lett. B}\ }\textbf {\bibinfo {volume} {422}},\ \bibinfo
  {pages} {247} (\bibinfo {year} {1998})},\ \Eprint
  {http://arxiv.org/abs/hep-ph/9711395} {arXiv:hep-ph/9711395} \BibitemShut
  {NoStop}%
\bibitem [{\citenamefont {Rapp}\ \emph {et~al.}(1998)\citenamefont {Rapp},
  \citenamefont {Sch{\"a}fer}, \citenamefont {Shuryak},\ and\ \citenamefont
  {Velkovsky}}]{Rapp:1997zu}%
  \BibitemOpen
  \bibfield  {author} {\bibinfo {author} {\bibfnamefont {R.}~\bibnamefont
  {Rapp}}, \bibinfo {author} {\bibfnamefont {T.}~\bibnamefont {Sch{\"a}fer}},
  \bibinfo {author} {\bibfnamefont {E.~V.}\ \bibnamefont {Shuryak}}, \ and\
  \bibinfo {author} {\bibfnamefont {M.}~\bibnamefont {Velkovsky}},\ }\href
  {\doibase 10.1103/PhysRevLett.81.53} {\bibfield  {journal} {\bibinfo
  {journal} {Phys. Rev. Lett.}\ }\textbf {\bibinfo {volume} {81}},\ \bibinfo
  {pages} {53} (\bibinfo {year} {1998})},\ \Eprint
  {http://arxiv.org/abs/hep-ph/9711396} {arXiv:hep-ph/9711396} \BibitemShut
  {NoStop}%
\bibitem [{\citenamefont {Eto}\ \emph {et~al.}(2014)\citenamefont {Eto},
  \citenamefont {Hirono}, \citenamefont {Nitta},\ and\ \citenamefont
  {Yasui}}]{Eto:2013hoa}%
  \BibitemOpen
  \bibfield  {author} {\bibinfo {author} {\bibfnamefont {M.}~\bibnamefont
  {Eto}}, \bibinfo {author} {\bibfnamefont {Y.}~\bibnamefont {Hirono}},
  \bibinfo {author} {\bibfnamefont {M.}~\bibnamefont {Nitta}}, \ and\ \bibinfo
  {author} {\bibfnamefont {S.}~\bibnamefont {Yasui}},\ }\href {\doibase
  10.1093/ptep/ptt095} {\bibfield  {journal} {\bibinfo  {journal} {PTEP}\
  }\textbf {\bibinfo {volume} {2014}},\ \bibinfo {pages} {012D01} (\bibinfo
  {year} {2014})},\ \Eprint {http://arxiv.org/abs/1308.1535} {arXiv:1308.1535
  [hep-ph]} \BibitemShut {NoStop}%
\bibitem [{\citenamefont {Forbes}\ and\ \citenamefont
  {Zhitnitsky}(2002)}]{Forbes:2001gj}%
  \BibitemOpen
  \bibfield  {author} {\bibinfo {author} {\bibfnamefont {M.~M.}\ \bibnamefont
  {Forbes}}\ and\ \bibinfo {author} {\bibfnamefont {A.~R.}\ \bibnamefont
  {Zhitnitsky}},\ }\href {\doibase 10.1103/PhysRevD.65.085009} {\bibfield
  {journal} {\bibinfo  {journal} {Phys. Rev. D}\ }\textbf {\bibinfo {volume}
  {65}},\ \bibinfo {pages} {085009} (\bibinfo {year} {2002})},\ \Eprint
  {http://arxiv.org/abs/hep-ph/0109173} {arXiv:hep-ph/0109173} \BibitemShut
  {NoStop}%
\bibitem [{\citenamefont {Iida}\ and\ \citenamefont
  {Baym}(2002)}]{Iida:2002ev}%
  \BibitemOpen
  \bibfield  {author} {\bibinfo {author} {\bibfnamefont {K.}~\bibnamefont
  {Iida}}\ and\ \bibinfo {author} {\bibfnamefont {G.}~\bibnamefont {Baym}},\
  }\href {\doibase 10.1103/PhysRevD.66.014015} {\bibfield  {journal} {\bibinfo
  {journal} {Phys. Rev. D}\ }\textbf {\bibinfo {volume} {66}},\ \bibinfo
  {pages} {014015} (\bibinfo {year} {2002})},\ \Eprint
  {http://arxiv.org/abs/hep-ph/0204124} {arXiv:hep-ph/0204124} \BibitemShut
  {NoStop}%
\bibitem [{\citenamefont {Vafa}\ and\ \citenamefont
  {Witten}(1984)}]{Vafa:1983tf}%
  \BibitemOpen
  \bibfield  {author} {\bibinfo {author} {\bibfnamefont {C.}~\bibnamefont
  {Vafa}}\ and\ \bibinfo {author} {\bibfnamefont {E.}~\bibnamefont {Witten}},\
  }\href {\doibase 10.1016/0550-3213(84)90230-X} {\bibfield  {journal}
  {\bibinfo  {journal} {Nucl. Phys. B}\ }\textbf {\bibinfo {volume} {234}},\
  \bibinfo {pages} {173} (\bibinfo {year} {1984})}\BibitemShut {NoStop}%
\bibitem [{\citenamefont {Balachandran}\ \emph {et~al.}(2006)\citenamefont
  {Balachandran}, \citenamefont {Digal},\ and\ \citenamefont
  {Matsuura}}]{Balachandran:2005ev}%
  \BibitemOpen
  \bibfield  {author} {\bibinfo {author} {\bibfnamefont {A.}~\bibnamefont
  {Balachandran}}, \bibinfo {author} {\bibfnamefont {S.}~\bibnamefont {Digal}},
  \ and\ \bibinfo {author} {\bibfnamefont {T.}~\bibnamefont {Matsuura}},\
  }\href {\doibase 10.1103/PhysRevD.73.074009} {\bibfield  {journal} {\bibinfo
  {journal} {Phys. Rev. D}\ }\textbf {\bibinfo {volume} {73}},\ \bibinfo
  {pages} {074009} (\bibinfo {year} {2006})},\ \Eprint
  {http://arxiv.org/abs/hep-ph/0509276} {arXiv:hep-ph/0509276} \BibitemShut
  {NoStop}%
\bibitem [{\citenamefont {Nakano}\ \emph
  {et~al.}(2008{\natexlab{a}})\citenamefont {Nakano}, \citenamefont {Nitta},\
  and\ \citenamefont {Matsuura}}]{Nakano:2007dr}%
  \BibitemOpen
  \bibfield  {author} {\bibinfo {author} {\bibfnamefont {E.}~\bibnamefont
  {Nakano}}, \bibinfo {author} {\bibfnamefont {M.}~\bibnamefont {Nitta}}, \
  and\ \bibinfo {author} {\bibfnamefont {T.}~\bibnamefont {Matsuura}},\ }\href
  {\doibase 10.1103/PhysRevD.78.045002} {\bibfield  {journal} {\bibinfo
  {journal} {Phys. Rev. D}\ }\textbf {\bibinfo {volume} {78}},\ \bibinfo
  {pages} {045002} (\bibinfo {year} {2008}{\natexlab{a}})},\ \Eprint
  {http://arxiv.org/abs/0708.4096} {arXiv:0708.4096 [hep-ph]} \BibitemShut
  {NoStop}%
\bibitem [{\citenamefont {Nakano}\ \emph
  {et~al.}(2008{\natexlab{b}})\citenamefont {Nakano}, \citenamefont {Nitta},\
  and\ \citenamefont {Matsuura}}]{Nakano:2008dc}%
  \BibitemOpen
  \bibfield  {author} {\bibinfo {author} {\bibfnamefont {E.}~\bibnamefont
  {Nakano}}, \bibinfo {author} {\bibfnamefont {M.}~\bibnamefont {Nitta}}, \
  and\ \bibinfo {author} {\bibfnamefont {T.}~\bibnamefont {Matsuura}},\ }\href
  {\doibase 10.1143/PTPS.174.254} {\bibfield  {journal} {\bibinfo  {journal}
  {Prog. Theor. Phys. Suppl.}\ }\textbf {\bibinfo {volume} {174}},\ \bibinfo
  {pages} {254} (\bibinfo {year} {2008}{\natexlab{b}})},\ \Eprint
  {http://arxiv.org/abs/0805.4539} {arXiv:0805.4539 [hep-ph]} \BibitemShut
  {NoStop}%
\bibitem [{\citenamefont {Eto}\ and\ \citenamefont {Nitta}(2009)}]{Eto:2009kg}%
  \BibitemOpen
  \bibfield  {author} {\bibinfo {author} {\bibfnamefont {M.}~\bibnamefont
  {Eto}}\ and\ \bibinfo {author} {\bibfnamefont {M.}~\bibnamefont {Nitta}},\
  }\href {\doibase 10.1103/PhysRevD.80.125007} {\bibfield  {journal} {\bibinfo
  {journal} {Phys. Rev. D}\ }\textbf {\bibinfo {volume} {80}},\ \bibinfo
  {pages} {125007} (\bibinfo {year} {2009})},\ \Eprint
  {http://arxiv.org/abs/0907.1278} {arXiv:0907.1278 [hep-ph]} \BibitemShut
  {NoStop}%
\bibitem [{\citenamefont {Sch\"afer}\ and\ \citenamefont
  {Wilczek}(1999)}]{Schafer:1998ef}%
  \BibitemOpen
  \bibfield  {author} {\bibinfo {author} {\bibfnamefont {T.}~\bibnamefont
  {Sch\"afer}}\ and\ \bibinfo {author} {\bibfnamefont {F.}~\bibnamefont
  {Wilczek}},\ }\href {\doibase 10.1103/PhysRevLett.82.3956} {\bibfield
  {journal} {\bibinfo  {journal} {Phys. Rev. Lett.}\ }\textbf {\bibinfo
  {volume} {82}},\ \bibinfo {pages} {3956} (\bibinfo {year} {1999})},\ \Eprint
  {http://arxiv.org/abs/hep-ph/9811473} {arXiv:hep-ph/9811473} \BibitemShut
  {NoStop}%
\bibitem [{\citenamefont {Alford}\ \emph
  {et~al.}(1999{\natexlab{b}})\citenamefont {Alford}, \citenamefont {Berges},\
  and\ \citenamefont {Rajagopal}}]{Alford:1999pa}%
  \BibitemOpen
  \bibfield  {author} {\bibinfo {author} {\bibfnamefont {M.~G.}\ \bibnamefont
  {Alford}}, \bibinfo {author} {\bibfnamefont {J.}~\bibnamefont {Berges}}, \
  and\ \bibinfo {author} {\bibfnamefont {K.}~\bibnamefont {Rajagopal}},\ }\href
  {\doibase 10.1016/S0550-3213(99)00410-1} {\bibfield  {journal} {\bibinfo
  {journal} {Nucl. Phys.}\ }\textbf {\bibinfo {volume} {B558}},\ \bibinfo
  {pages} {219} (\bibinfo {year} {1999}{\natexlab{b}})},\ \Eprint
  {http://arxiv.org/abs/hep-ph/9903502} {arXiv:hep-ph/9903502 [hep-ph]}
  \BibitemShut {NoStop}%
\bibitem [{\citenamefont {Fukushima}(2004)}]{Fukushima:2004bj}%
  \BibitemOpen
  \bibfield  {author} {\bibinfo {author} {\bibfnamefont {K.}~\bibnamefont
  {Fukushima}},\ }\href {\doibase 10.1103/PhysRevD.70.094014} {\bibfield
  {journal} {\bibinfo  {journal} {Phys. Rev. D}\ }\textbf {\bibinfo {volume}
  {70}},\ \bibinfo {pages} {094014} (\bibinfo {year} {2004})},\ \Eprint
  {http://arxiv.org/abs/hep-ph/0403091} {arXiv:hep-ph/0403091} \BibitemShut
  {NoStop}%
\bibitem [{\citenamefont {Hatsuda}\ \emph {et~al.}(2006)\citenamefont
  {Hatsuda}, \citenamefont {Tachibana}, \citenamefont {Yamamoto},\ and\
  \citenamefont {Baym}}]{Hatsuda:2006ps}%
  \BibitemOpen
  \bibfield  {author} {\bibinfo {author} {\bibfnamefont {T.}~\bibnamefont
  {Hatsuda}}, \bibinfo {author} {\bibfnamefont {M.}~\bibnamefont {Tachibana}},
  \bibinfo {author} {\bibfnamefont {N.}~\bibnamefont {Yamamoto}}, \ and\
  \bibinfo {author} {\bibfnamefont {G.}~\bibnamefont {Baym}},\ }\href {\doibase
  10.1103/PhysRevLett.97.122001} {\bibfield  {journal} {\bibinfo  {journal}
  {Phys. Rev. Lett.}\ }\textbf {\bibinfo {volume} {97}},\ \bibinfo {pages}
  {122001} (\bibinfo {year} {2006})},\ \Eprint
  {http://arxiv.org/abs/hep-ph/0605018} {arXiv:hep-ph/0605018} \BibitemShut
  {NoStop}%
\bibitem [{\citenamefont {Yamamoto}\ \emph {et~al.}(2007)\citenamefont
  {Yamamoto}, \citenamefont {Tachibana}, \citenamefont {Hatsuda},\ and\
  \citenamefont {Baym}}]{Yamamoto:2007ah}%
  \BibitemOpen
  \bibfield  {author} {\bibinfo {author} {\bibfnamefont {N.}~\bibnamefont
  {Yamamoto}}, \bibinfo {author} {\bibfnamefont {M.}~\bibnamefont {Tachibana}},
  \bibinfo {author} {\bibfnamefont {T.}~\bibnamefont {Hatsuda}}, \ and\
  \bibinfo {author} {\bibfnamefont {G.}~\bibnamefont {Baym}},\ }\href {\doibase
  10.1103/PhysRevD.76.074001} {\bibfield  {journal} {\bibinfo  {journal} {Phys.
  Rev.}\ }\textbf {\bibinfo {volume} {D76}},\ \bibinfo {pages} {074001}
  (\bibinfo {year} {2007})},\ \Eprint {http://arxiv.org/abs/0704.2654}
  {arXiv:0704.2654 [hep-ph]} \BibitemShut {NoStop}%
\bibitem [{\citenamefont {Hatsuda}\ \emph {et~al.}(2008)\citenamefont
  {Hatsuda}, \citenamefont {Tachibana},\ and\ \citenamefont
  {Yamamoto}}]{Hatsuda:2008is}%
  \BibitemOpen
  \bibfield  {author} {\bibinfo {author} {\bibfnamefont {T.}~\bibnamefont
  {Hatsuda}}, \bibinfo {author} {\bibfnamefont {M.}~\bibnamefont {Tachibana}},
  \ and\ \bibinfo {author} {\bibfnamefont {N.}~\bibnamefont {Yamamoto}},\
  }\href {\doibase 10.1103/PhysRevD.78.011501} {\bibfield  {journal} {\bibinfo
  {journal} {Phys. Rev.}\ }\textbf {\bibinfo {volume} {D78}},\ \bibinfo {pages}
  {011501} (\bibinfo {year} {2008})},\ \Eprint {http://arxiv.org/abs/0802.4143}
  {arXiv:0802.4143 [hep-ph]} \BibitemShut {NoStop}%
\bibitem [{\citenamefont {Schmitt}\ \emph {et~al.}(2011)\citenamefont
  {Schmitt}, \citenamefont {Stetina},\ and\ \citenamefont
  {Tachibana}}]{Schmitt:2010pf}%
  \BibitemOpen
  \bibfield  {author} {\bibinfo {author} {\bibfnamefont {A.}~\bibnamefont
  {Schmitt}}, \bibinfo {author} {\bibfnamefont {S.}~\bibnamefont {Stetina}}, \
  and\ \bibinfo {author} {\bibfnamefont {M.}~\bibnamefont {Tachibana}},\ }\href
  {\doibase 10.1103/PhysRevD.83.045008} {\bibfield  {journal} {\bibinfo
  {journal} {Phys. Rev. D}\ }\textbf {\bibinfo {volume} {83}},\ \bibinfo
  {pages} {045008} (\bibinfo {year} {2011})},\ \Eprint
  {http://arxiv.org/abs/1010.4243} {arXiv:1010.4243 [hep-ph]} \BibitemShut
  {NoStop}%
\bibitem [{\citenamefont {Alford}\ \emph {et~al.}(2019)\citenamefont {Alford},
  \citenamefont {Baym}, \citenamefont {Fukushima}, \citenamefont {Hatsuda},\
  and\ \citenamefont {Tachibana}}]{Alford:2018mqj}%
  \BibitemOpen
  \bibfield  {author} {\bibinfo {author} {\bibfnamefont {M.~G.}\ \bibnamefont
  {Alford}}, \bibinfo {author} {\bibfnamefont {G.}~\bibnamefont {Baym}},
  \bibinfo {author} {\bibfnamefont {K.}~\bibnamefont {Fukushima}}, \bibinfo
  {author} {\bibfnamefont {T.}~\bibnamefont {Hatsuda}}, \ and\ \bibinfo
  {author} {\bibfnamefont {M.}~\bibnamefont {Tachibana}},\ }\href {\doibase
  10.1103/PhysRevD.99.036004} {\bibfield  {journal} {\bibinfo  {journal} {Phys.
  Rev. D}\ }\textbf {\bibinfo {volume} {99}},\ \bibinfo {pages} {036004}
  (\bibinfo {year} {2019})},\ \Eprint {http://arxiv.org/abs/1803.05115}
  {arXiv:1803.05115 [hep-ph]} \BibitemShut {NoStop}%
\bibitem [{\citenamefont {Chatterjee}\ \emph
  {et~al.}(2019{\natexlab{a}})\citenamefont {Chatterjee}, \citenamefont
  {Nitta},\ and\ \citenamefont {Yasui}}]{Chatterjee:2018nxe}%
  \BibitemOpen
  \bibfield  {author} {\bibinfo {author} {\bibfnamefont {C.}~\bibnamefont
  {Chatterjee}}, \bibinfo {author} {\bibfnamefont {M.}~\bibnamefont {Nitta}}, \
  and\ \bibinfo {author} {\bibfnamefont {S.}~\bibnamefont {Yasui}},\ }\href
  {\doibase 10.1103/PhysRevD.99.034001} {\bibfield  {journal} {\bibinfo
  {journal} {Phys. Rev. D}\ }\textbf {\bibinfo {volume} {99}},\ \bibinfo
  {pages} {034001} (\bibinfo {year} {2019}{\natexlab{a}})},\ \Eprint
  {http://arxiv.org/abs/1806.09291} {arXiv:1806.09291 [hep-ph]} \BibitemShut
  {NoStop}%
\bibitem [{\citenamefont {Chatterjee}\ \emph
  {et~al.}(2019{\natexlab{b}})\citenamefont {Chatterjee}, \citenamefont
  {Nitta},\ and\ \citenamefont {Yasui}}]{Chatterjee:2019tbz}%
  \BibitemOpen
  \bibfield  {author} {\bibinfo {author} {\bibfnamefont {C.}~\bibnamefont
  {Chatterjee}}, \bibinfo {author} {\bibfnamefont {M.}~\bibnamefont {Nitta}}, \
  and\ \bibinfo {author} {\bibfnamefont {S.}~\bibnamefont {Yasui}},\ }\href
  {\doibase 10.7566/JPSCP.26.024030} {\bibfield  {journal} {\bibinfo  {journal}
  {JPS Conf. Proc.}\ }\textbf {\bibinfo {volume} {26}},\ \bibinfo {pages}
  {024030} (\bibinfo {year} {2019}{\natexlab{b}})},\ \Eprint
  {http://arxiv.org/abs/1902.00156} {arXiv:1902.00156 [hep-ph]} \BibitemShut
  {NoStop}%
\bibitem [{\citenamefont {Cherman}\ \emph {et~al.}(2019)\citenamefont
  {Cherman}, \citenamefont {Sen},\ and\ \citenamefont
  {Yaffe}}]{Cherman:2018jir}%
  \BibitemOpen
  \bibfield  {author} {\bibinfo {author} {\bibfnamefont {A.}~\bibnamefont
  {Cherman}}, \bibinfo {author} {\bibfnamefont {S.}~\bibnamefont {Sen}}, \ and\
  \bibinfo {author} {\bibfnamefont {L.~G.}\ \bibnamefont {Yaffe}},\ }\href
  {\doibase 10.1103/PhysRevD.100.034015} {\bibfield  {journal} {\bibinfo
  {journal} {Phys. Rev. D}\ }\textbf {\bibinfo {volume} {100}},\ \bibinfo
  {pages} {034015} (\bibinfo {year} {2019})},\ \Eprint
  {http://arxiv.org/abs/1808.04827} {arXiv:1808.04827 [hep-th]} \BibitemShut
  {NoStop}%
\bibitem [{\citenamefont {Hirono}\ and\ \citenamefont
  {Tanizaki}(2019{\natexlab{a}})}]{Hirono:2018fjr}%
  \BibitemOpen
  \bibfield  {author} {\bibinfo {author} {\bibfnamefont {Y.}~\bibnamefont
  {Hirono}}\ and\ \bibinfo {author} {\bibfnamefont {Y.}~\bibnamefont
  {Tanizaki}},\ }\href {\doibase 10.1103/PhysRevLett.122.212001} {\bibfield
  {journal} {\bibinfo  {journal} {Phys. Rev. Lett.}\ }\textbf {\bibinfo
  {volume} {122}},\ \bibinfo {pages} {212001} (\bibinfo {year}
  {2019}{\natexlab{a}})},\ \Eprint {http://arxiv.org/abs/1811.10608}
  {arXiv:1811.10608 [hep-th]} \BibitemShut {NoStop}%
\bibitem [{\citenamefont {Hirono}\ and\ \citenamefont
  {Tanizaki}(2019{\natexlab{b}})}]{Hirono:2019oup}%
  \BibitemOpen
  \bibfield  {author} {\bibinfo {author} {\bibfnamefont {Y.}~\bibnamefont
  {Hirono}}\ and\ \bibinfo {author} {\bibfnamefont {Y.}~\bibnamefont
  {Tanizaki}},\ }\href {\doibase 10.1007/JHEP07(2019)062} {\bibfield  {journal}
  {\bibinfo  {journal} {JHEP}\ }\textbf {\bibinfo {volume} {07}},\ \bibinfo
  {pages} {062} (\bibinfo {year} {2019}{\natexlab{b}})},\ \Eprint
  {http://arxiv.org/abs/1904.08570} {arXiv:1904.08570 [hep-th]} \BibitemShut
  {NoStop}%
\bibitem [{\citenamefont {Cherman}\ \emph {et~al.}(2020)\citenamefont
  {Cherman}, \citenamefont {Jacobson}, \citenamefont {Sen},\ and\ \citenamefont
  {Yaffe}}]{Cherman:2020hbe}%
  \BibitemOpen
  \bibfield  {author} {\bibinfo {author} {\bibfnamefont {A.}~\bibnamefont
  {Cherman}}, \bibinfo {author} {\bibfnamefont {T.}~\bibnamefont {Jacobson}},
  \bibinfo {author} {\bibfnamefont {S.}~\bibnamefont {Sen}}, \ and\ \bibinfo
  {author} {\bibfnamefont {L.~G.}\ \bibnamefont {Yaffe}},\ }\href@noop {} {\
  (\bibinfo {year} {2020})},\ \Eprint {http://arxiv.org/abs/2007.08539}
  {arXiv:2007.08539 [hep-th]} \BibitemShut {NoStop}%
\bibitem [{\citenamefont {Alford}\ and\ \citenamefont
  {Sedrakian}(2010)}]{Alford:2010qf}%
  \BibitemOpen
  \bibfield  {author} {\bibinfo {author} {\bibfnamefont {M.~G.}\ \bibnamefont
  {Alford}}\ and\ \bibinfo {author} {\bibfnamefont {A.}~\bibnamefont
  {Sedrakian}},\ }\href {\doibase 10.1088/0954-3899/37/7/075202} {\bibfield
  {journal} {\bibinfo  {journal} {J. Phys. G}\ }\textbf {\bibinfo {volume}
  {37}},\ \bibinfo {pages} {075202} (\bibinfo {year} {2010})},\ \Eprint
  {http://arxiv.org/abs/1001.3346} {arXiv:1001.3346 [astro-ph.SR]} \BibitemShut
  {NoStop}%
\bibitem [{\citenamefont {Fujimoto}\ \emph {et~al.}(2020)\citenamefont
  {Fujimoto}, \citenamefont {Fukushima},\ and\ \citenamefont
  {Weise}}]{Fujimoto:2019sxg}%
  \BibitemOpen
  \bibfield  {author} {\bibinfo {author} {\bibfnamefont {Y.}~\bibnamefont
  {Fujimoto}}, \bibinfo {author} {\bibfnamefont {K.}~\bibnamefont {Fukushima}},
  \ and\ \bibinfo {author} {\bibfnamefont {W.}~\bibnamefont {Weise}},\ }\href
  {\doibase 10.1103/PhysRevD.101.094009} {\bibfield  {journal} {\bibinfo
  {journal} {Phys. Rev. D}\ }\textbf {\bibinfo {volume} {101}},\ \bibinfo
  {pages} {094009} (\bibinfo {year} {2020})},\ \Eprint
  {http://arxiv.org/abs/1908.09360} {arXiv:1908.09360 [hep-ph]} \BibitemShut
  {NoStop}%
\bibitem [{\citenamefont {Fujimoto}(2020)}]{Fujimoto:2020cho}%
  \BibitemOpen
  \bibfield  {author} {\bibinfo {author} {\bibfnamefont {Y.}~\bibnamefont
  {Fujimoto}}\ }(\bibinfo {year} {2020})\ \Eprint
  {http://arxiv.org/abs/2002.08073} {arXiv:2002.08073 [hep-ph]} \BibitemShut
  {NoStop}%
\bibitem [{\citenamefont {Annala}\ \emph {et~al.}(2020)\citenamefont {Annala},
  \citenamefont {Gorda}, \citenamefont {Kurkela}, \citenamefont {N\"attil\"a},\
  and\ \citenamefont {Vuorinen}}]{Annala:2019puf}%
  \BibitemOpen
  \bibfield  {author} {\bibinfo {author} {\bibfnamefont {E.}~\bibnamefont
  {Annala}}, \bibinfo {author} {\bibfnamefont {T.}~\bibnamefont {Gorda}},
  \bibinfo {author} {\bibfnamefont {A.}~\bibnamefont {Kurkela}}, \bibinfo
  {author} {\bibfnamefont {J.}~\bibnamefont {N\"attil\"a}}, \ and\ \bibinfo
  {author} {\bibfnamefont {A.}~\bibnamefont {Vuorinen}},\ }\href {\doibase
  10.1038/s41567-020-0914-9} {\bibfield  {journal} {\bibinfo  {journal} {Nature
  Phys.}\ } (\bibinfo {year} {2020}),\ 10.1038/s41567-020-0914-9},\ \Eprint
  {http://arxiv.org/abs/1903.09121} {arXiv:1903.09121 [astro-ph.HE]}
  \BibitemShut {NoStop}%
\bibitem [{\citenamefont {Masuda}\ \emph
  {et~al.}(2013{\natexlab{a}})\citenamefont {Masuda}, \citenamefont {Hatsuda},\
  and\ \citenamefont {Takatsuka}}]{Masuda:2012kf}%
  \BibitemOpen
  \bibfield  {author} {\bibinfo {author} {\bibfnamefont {K.}~\bibnamefont
  {Masuda}}, \bibinfo {author} {\bibfnamefont {T.}~\bibnamefont {Hatsuda}}, \
  and\ \bibinfo {author} {\bibfnamefont {T.}~\bibnamefont {Takatsuka}},\ }\href
  {\doibase 10.1088/0004-637X/764/1/12} {\bibfield  {journal} {\bibinfo
  {journal} {Astrophys. J.}\ }\textbf {\bibinfo {volume} {764}},\ \bibinfo
  {pages} {12} (\bibinfo {year} {2013}{\natexlab{a}})},\ \Eprint
  {http://arxiv.org/abs/1205.3621} {arXiv:1205.3621 [nucl-th]} \BibitemShut
  {NoStop}%
\bibitem [{\citenamefont {Masuda}\ \emph
  {et~al.}(2013{\natexlab{b}})\citenamefont {Masuda}, \citenamefont {Hatsuda},\
  and\ \citenamefont {Takatsuka}}]{Masuda:2012ed}%
  \BibitemOpen
  \bibfield  {author} {\bibinfo {author} {\bibfnamefont {K.}~\bibnamefont
  {Masuda}}, \bibinfo {author} {\bibfnamefont {T.}~\bibnamefont {Hatsuda}}, \
  and\ \bibinfo {author} {\bibfnamefont {T.}~\bibnamefont {Takatsuka}},\ }\href
  {\doibase 10.1093/ptep/ptt045} {\bibfield  {journal} {\bibinfo  {journal}
  {PTEP}\ }\textbf {\bibinfo {volume} {2013}},\ \bibinfo {pages} {073D01}
  (\bibinfo {year} {2013}{\natexlab{b}})},\ \Eprint
  {http://arxiv.org/abs/1212.6803} {arXiv:1212.6803 [nucl-th]} \BibitemShut
  {NoStop}%
\bibitem [{\citenamefont {Kojo}\ \emph {et~al.}(2015)\citenamefont {Kojo},
  \citenamefont {Powell}, \citenamefont {Song},\ and\ \citenamefont
  {Baym}}]{Kojo:2014rca}%
  \BibitemOpen
  \bibfield  {author} {\bibinfo {author} {\bibfnamefont {T.}~\bibnamefont
  {Kojo}}, \bibinfo {author} {\bibfnamefont {P.~D.}\ \bibnamefont {Powell}},
  \bibinfo {author} {\bibfnamefont {Y.}~\bibnamefont {Song}}, \ and\ \bibinfo
  {author} {\bibfnamefont {G.}~\bibnamefont {Baym}},\ }\href {\doibase
  10.1103/PhysRevD.91.045003} {\bibfield  {journal} {\bibinfo  {journal} {Phys.
  Rev. D}\ }\textbf {\bibinfo {volume} {91}},\ \bibinfo {pages} {045003}
  (\bibinfo {year} {2015})},\ \Eprint {http://arxiv.org/abs/1412.1108}
  {arXiv:1412.1108 [hep-ph]} \BibitemShut {NoStop}%
\bibitem [{\citenamefont {Baym}\ \emph {et~al.}(2018)\citenamefont {Baym},
  \citenamefont {Hatsuda}, \citenamefont {Kojo}, \citenamefont {Powell},
  \citenamefont {Song},\ and\ \citenamefont {Takatsuka}}]{Baym:2017whm}%
  \BibitemOpen
  \bibfield  {author} {\bibinfo {author} {\bibfnamefont {G.}~\bibnamefont
  {Baym}}, \bibinfo {author} {\bibfnamefont {T.}~\bibnamefont {Hatsuda}},
  \bibinfo {author} {\bibfnamefont {T.}~\bibnamefont {Kojo}}, \bibinfo {author}
  {\bibfnamefont {P.~D.}\ \bibnamefont {Powell}}, \bibinfo {author}
  {\bibfnamefont {Y.}~\bibnamefont {Song}}, \ and\ \bibinfo {author}
  {\bibfnamefont {T.}~\bibnamefont {Takatsuka}},\ }\href {\doibase
  10.1088/1361-6633/aaae14} {\bibfield  {journal} {\bibinfo  {journal} {Rept.
  Prog. Phys.}\ }\textbf {\bibinfo {volume} {81}},\ \bibinfo {pages} {056902}
  (\bibinfo {year} {2018})},\ \Eprint {http://arxiv.org/abs/1707.04966}
  {arXiv:1707.04966 [astro-ph.HE]} \BibitemShut {NoStop}%
\bibitem [{\citenamefont {Demorest}\ \emph {et~al.}(2010)\citenamefont
  {Demorest}, \citenamefont {Pennucci}, \citenamefont {Ransom}, \citenamefont
  {Roberts},\ and\ \citenamefont {Hessels}}]{Demorest:2010bx}%
  \BibitemOpen
  \bibfield  {author} {\bibinfo {author} {\bibfnamefont {P.}~\bibnamefont
  {Demorest}}, \bibinfo {author} {\bibfnamefont {T.}~\bibnamefont {Pennucci}},
  \bibinfo {author} {\bibfnamefont {S.}~\bibnamefont {Ransom}}, \bibinfo
  {author} {\bibfnamefont {M.}~\bibnamefont {Roberts}}, \ and\ \bibinfo
  {author} {\bibfnamefont {J.}~\bibnamefont {Hessels}},\ }\href {\doibase
  10.1038/nature09466} {\bibfield  {journal} {\bibinfo  {journal} {Nature}\
  }\textbf {\bibinfo {volume} {467}},\ \bibinfo {pages} {1081} (\bibinfo {year}
  {2010})},\ \Eprint {http://arxiv.org/abs/1010.5788} {arXiv:1010.5788
  [astro-ph.HE]} \BibitemShut {NoStop}%
\bibitem [{\citenamefont {Fonseca}\ \emph {et~al.}(2016)\citenamefont {Fonseca}
  \emph {et~al.}}]{Fonseca:2016tux}%
  \BibitemOpen
  \bibfield  {author} {\bibinfo {author} {\bibfnamefont {E.}~\bibnamefont
  {Fonseca}} \emph {et~al.},\ }\href {\doibase 10.3847/0004-637X/832/2/167}
  {\bibfield  {journal} {\bibinfo  {journal} {Astrophys. J.}\ }\textbf
  {\bibinfo {volume} {832}},\ \bibinfo {pages} {167} (\bibinfo {year}
  {2016})},\ \Eprint {http://arxiv.org/abs/1603.00545} {arXiv:1603.00545
  [astro-ph.HE]} \BibitemShut {NoStop}%
\bibitem [{\citenamefont {Antoniadis}\ \emph {et~al.}(2013)\citenamefont
  {Antoniadis} \emph {et~al.}}]{Antoniadis:2013pzd}%
  \BibitemOpen
  \bibfield  {author} {\bibinfo {author} {\bibfnamefont {J.}~\bibnamefont
  {Antoniadis}} \emph {et~al.},\ }\href {\doibase 10.1126/science.1233232}
  {\bibfield  {journal} {\bibinfo  {journal} {Science}\ }\textbf {\bibinfo
  {volume} {340}},\ \bibinfo {pages} {6131} (\bibinfo {year} {2013})},\ \Eprint
  {http://arxiv.org/abs/1304.6875} {arXiv:1304.6875 [astro-ph.HE]} \BibitemShut
  {NoStop}%
\bibitem [{\citenamefont {Cromartie}\ \emph {et~al.}(2019)\citenamefont
  {Cromartie} \emph {et~al.}}]{Cromartie:2019kug}%
  \BibitemOpen
  \bibfield  {author} {\bibinfo {author} {\bibfnamefont {H.~T.}\ \bibnamefont
  {Cromartie}} \emph {et~al.},\ }\href {\doibase 10.1038/s41550-019-0880-2}
  {\bibfield  {journal} {\bibinfo  {journal} {Nature Astron.}\ }\textbf
  {\bibinfo {volume} {4}},\ \bibinfo {pages} {72} (\bibinfo {year} {2019})},\
  \Eprint {http://arxiv.org/abs/1904.06759} {arXiv:1904.06759 [astro-ph.HE]}
  \BibitemShut {NoStop}%
\bibitem [{\citenamefont {Hoffberg}\ \emph {et~al.}(1970)\citenamefont
  {Hoffberg}, \citenamefont {Glassgold}, \citenamefont {Richardson},\ and\
  \citenamefont {Ruderman}}]{Hoffberg:1970vqj}%
  \BibitemOpen
  \bibfield  {author} {\bibinfo {author} {\bibfnamefont {M.}~\bibnamefont
  {Hoffberg}}, \bibinfo {author} {\bibfnamefont {A.}~\bibnamefont {Glassgold}},
  \bibinfo {author} {\bibfnamefont {R.}~\bibnamefont {Richardson}}, \ and\
  \bibinfo {author} {\bibfnamefont {M.}~\bibnamefont {Ruderman}},\ }\href
  {\doibase 10.1103/PhysRevLett.24.775} {\bibfield  {journal} {\bibinfo
  {journal} {Phys. Rev. Lett.}\ }\textbf {\bibinfo {volume} {24}},\ \bibinfo
  {pages} {775} (\bibinfo {year} {1970})}\BibitemShut {NoStop}%
\bibitem [{\citenamefont {Tamagaki}(1970)}]{Tamagaki:1970ptp}%
  \BibitemOpen
  \bibfield  {author} {\bibinfo {author} {\bibfnamefont {R.}~\bibnamefont
  {Tamagaki}},\ }\href {\doibase 10.1143/PTP.44.905} {\bibfield  {journal}
  {\bibinfo  {journal} {Prog. Theor. Phys.}\ }\textbf {\bibinfo {volume}
  {44}},\ \bibinfo {pages} {905} (\bibinfo {year} {1970})}\BibitemShut
  {NoStop}%
\bibitem [{\citenamefont {Takatsuka}\ and\ \citenamefont
  {Tamagaki}(1971)}]{takatsukaPTP71}%
  \BibitemOpen
  \bibfield  {author} {\bibinfo {author} {\bibfnamefont {T.}~\bibnamefont
  {Takatsuka}}\ and\ \bibinfo {author} {\bibfnamefont {R.}~\bibnamefont
  {Tamagaki}},\ }\href {\doibase 10.1143/PTP.46.114} {\bibfield  {journal}
  {\bibinfo  {journal} {Prog. Theor. Phys.}\ }\textbf {\bibinfo {volume}
  {46}},\ \bibinfo {pages} {114} (\bibinfo {year} {1971})}\BibitemShut
  {NoStop}%
\bibitem [{\citenamefont {Takatsuka}(1972)}]{takatsukaPTP72}%
  \BibitemOpen
  \bibfield  {author} {\bibinfo {author} {\bibfnamefont {T.}~\bibnamefont
  {Takatsuka}},\ }\href {\doibase 10.1143/PTP.47.1062} {\bibfield  {journal}
  {\bibinfo  {journal} {Prog. Theor. Phys.}\ }\textbf {\bibinfo {volume}
  {47}},\ \bibinfo {pages} {1062} (\bibinfo {year} {1972})}\BibitemShut
  {NoStop}%
\bibitem [{\citenamefont {Richardson}(1972)}]{richardsonPRD72}%
  \BibitemOpen
  \bibfield  {author} {\bibinfo {author} {\bibfnamefont {R.~W.}\ \bibnamefont
  {Richardson}},\ }\href {\doibase 10.1103/PhysRevD.5.1883} {\bibfield
  {journal} {\bibinfo  {journal} {Phys. Rev. D}\ }\textbf {\bibinfo {volume}
  {5}},\ \bibinfo {pages} {1883} (\bibinfo {year} {1972})}\BibitemShut
  {NoStop}%
\bibitem [{\citenamefont {Sauls}\ and\ \citenamefont
  {Serene}(1978)}]{Sauls:1978lna}%
  \BibitemOpen
  \bibfield  {author} {\bibinfo {author} {\bibfnamefont {J.}~\bibnamefont
  {Sauls}}\ and\ \bibinfo {author} {\bibfnamefont {J.}~\bibnamefont {Serene}},\
  }\href {\doibase 10.1103/PhysRevD.17.1524} {\bibfield  {journal} {\bibinfo
  {journal} {Phys. Rev. D}\ }\textbf {\bibinfo {volume} {17}},\ \bibinfo
  {pages} {1524} (\bibinfo {year} {1978})}\BibitemShut {NoStop}%
\bibitem [{\citenamefont {Takatsuka}\ and\ \citenamefont
  {Tamagaki}(1993)}]{Takatsuka:1992ga}%
  \BibitemOpen
  \bibfield  {author} {\bibinfo {author} {\bibfnamefont {T.}~\bibnamefont
  {Takatsuka}}\ and\ \bibinfo {author} {\bibfnamefont {R.}~\bibnamefont
  {Tamagaki}},\ }\href {\doibase 10.1143/PTPS.112.27} {\bibfield  {journal}
  {\bibinfo  {journal} {Prog. Theor. Phys. Suppl.}\ }\textbf {\bibinfo {volume}
  {112}},\ \bibinfo {pages} {27} (\bibinfo {year} {1993})}\BibitemShut
  {NoStop}%
\bibitem [{\citenamefont {Mizushima}\ \emph {et~al.}(2017)\citenamefont
  {Mizushima}, \citenamefont {Masuda},\ and\ \citenamefont
  {Nitta}}]{Mizushima:2016fbn}%
  \BibitemOpen
  \bibfield  {author} {\bibinfo {author} {\bibfnamefont {T.}~\bibnamefont
  {Mizushima}}, \bibinfo {author} {\bibfnamefont {K.}~\bibnamefont {Masuda}}, \
  and\ \bibinfo {author} {\bibfnamefont {M.}~\bibnamefont {Nitta}},\ }\href
  {\doibase 10.1103/PhysRevB.95.140503} {\bibfield  {journal} {\bibinfo
  {journal} {Phys. Rev.}\ }\textbf {\bibinfo {volume} {B95}},\ \bibinfo {pages}
  {140503} (\bibinfo {year} {2017})},\ \Eprint
  {http://arxiv.org/abs/1607.07266} {arXiv:1607.07266 [cond-mat.supr-con]}
  \BibitemShut {NoStop}%
\bibitem [{\citenamefont {Yasui}\ \emph
  {et~al.}(2019{\natexlab{a}})\citenamefont {Yasui}, \citenamefont
  {Chatterjee},\ and\ \citenamefont {Nitta}}]{Yasui:2018tcr}%
  \BibitemOpen
  \bibfield  {author} {\bibinfo {author} {\bibfnamefont {S.}~\bibnamefont
  {Yasui}}, \bibinfo {author} {\bibfnamefont {C.}~\bibnamefont {Chatterjee}}, \
  and\ \bibinfo {author} {\bibfnamefont {M.}~\bibnamefont {Nitta}},\ }\href
  {\doibase 10.1103/PhysRevC.99.035213} {\bibfield  {journal} {\bibinfo
  {journal} {Phys. Rev. C}\ }\textbf {\bibinfo {volume} {99}},\ \bibinfo
  {pages} {035213} (\bibinfo {year} {2019}{\natexlab{a}})},\ \Eprint
  {http://arxiv.org/abs/1810.04901} {arXiv:1810.04901 [nucl-th]} \BibitemShut
  {NoStop}%
\bibitem [{\citenamefont {Yasui}\ \emph
  {et~al.}(2019{\natexlab{b}})\citenamefont {Yasui}, \citenamefont
  {Chatterjee}, \citenamefont {Kobayashi},\ and\ \citenamefont
  {Nitta}}]{Yasui:2019unp}%
  \BibitemOpen
  \bibfield  {author} {\bibinfo {author} {\bibfnamefont {S.}~\bibnamefont
  {Yasui}}, \bibinfo {author} {\bibfnamefont {C.}~\bibnamefont {Chatterjee}},
  \bibinfo {author} {\bibfnamefont {M.}~\bibnamefont {Kobayashi}}, \ and\
  \bibinfo {author} {\bibfnamefont {M.}~\bibnamefont {Nitta}},\ }\href
  {\doibase 10.1103/PhysRevC.100.025204} {\bibfield  {journal} {\bibinfo
  {journal} {Phys. Rev. C}\ }\textbf {\bibinfo {volume} {100}},\ \bibinfo
  {pages} {025204} (\bibinfo {year} {2019}{\natexlab{b}})},\ \Eprint
  {http://arxiv.org/abs/1904.11399} {arXiv:1904.11399 [nucl-th]} \BibitemShut
  {NoStop}%
\bibitem [{\citenamefont {Schwarz}(1982)}]{Schwarz:1982ec}%
  \BibitemOpen
  \bibfield  {author} {\bibinfo {author} {\bibfnamefont {A.~S.}\ \bibnamefont
  {Schwarz}},\ }\href {\doibase 10.1016/0550-3213(82)90190-0} {\bibfield
  {journal} {\bibinfo  {journal} {Nucl. Phys. B}\ }\textbf {\bibinfo {volume}
  {208}},\ \bibinfo {pages} {141} (\bibinfo {year} {1982})}\BibitemShut
  {NoStop}%
\bibitem [{\citenamefont {Alford}\ \emph {et~al.}(1990)\citenamefont {Alford},
  \citenamefont {Benson}, \citenamefont {Coleman}, \citenamefont
  {March-Russell},\ and\ \citenamefont {Wilczek}}]{Alford:1990mk}%
  \BibitemOpen
  \bibfield  {author} {\bibinfo {author} {\bibfnamefont {M.~G.}\ \bibnamefont
  {Alford}}, \bibinfo {author} {\bibfnamefont {K.}~\bibnamefont {Benson}},
  \bibinfo {author} {\bibfnamefont {S.~R.}\ \bibnamefont {Coleman}}, \bibinfo
  {author} {\bibfnamefont {J.}~\bibnamefont {March-Russell}}, \ and\ \bibinfo
  {author} {\bibfnamefont {F.}~\bibnamefont {Wilczek}},\ }\href {\doibase
  10.1103/PhysRevLett.64.1632} {\bibfield  {journal} {\bibinfo  {journal}
  {Phys. Rev. Lett.}\ }\textbf {\bibinfo {volume} {64}},\ \bibinfo {pages}
  {1632} (\bibinfo {year} {1990})},\ \bibinfo {note} {[Erratum: Phys.Rev.Lett.
  65, 668 (1990)]}\BibitemShut {NoStop}%
\bibitem [{\citenamefont {Alford}\ \emph {et~al.}(1991)\citenamefont {Alford},
  \citenamefont {Benson}, \citenamefont {Coleman}, \citenamefont
  {March-Russell},\ and\ \citenamefont {Wilczek}}]{Alford:1990ur}%
  \BibitemOpen
  \bibfield  {author} {\bibinfo {author} {\bibfnamefont {M.~G.}\ \bibnamefont
  {Alford}}, \bibinfo {author} {\bibfnamefont {K.}~\bibnamefont {Benson}},
  \bibinfo {author} {\bibfnamefont {S.~R.}\ \bibnamefont {Coleman}}, \bibinfo
  {author} {\bibfnamefont {J.}~\bibnamefont {March-Russell}}, \ and\ \bibinfo
  {author} {\bibfnamefont {F.}~\bibnamefont {Wilczek}},\ }\href {\doibase
  10.1016/0550-3213(91)90331-Q} {\bibfield  {journal} {\bibinfo  {journal}
  {Nucl. Phys. B}\ }\textbf {\bibinfo {volume} {349}},\ \bibinfo {pages} {414}
  (\bibinfo {year} {1991})}\BibitemShut {NoStop}%
\bibitem [{\citenamefont {Alford}\ \emph {et~al.}(1992)\citenamefont {Alford},
  \citenamefont {Lee}, \citenamefont {March-Russell},\ and\ \citenamefont
  {Preskill}}]{Alford:1992yx}%
  \BibitemOpen
  \bibfield  {author} {\bibinfo {author} {\bibfnamefont {M.~G.}\ \bibnamefont
  {Alford}}, \bibinfo {author} {\bibfnamefont {K.-M.}\ \bibnamefont {Lee}},
  \bibinfo {author} {\bibfnamefont {J.}~\bibnamefont {March-Russell}}, \ and\
  \bibinfo {author} {\bibfnamefont {J.}~\bibnamefont {Preskill}},\ }\href
  {\doibase 10.1016/0550-3213(92)90468-Q} {\bibfield  {journal} {\bibinfo
  {journal} {Nucl. Phys. B}\ }\textbf {\bibinfo {volume} {384}},\ \bibinfo
  {pages} {251} (\bibinfo {year} {1992})},\ \Eprint
  {http://arxiv.org/abs/hep-th/9112038} {arXiv:hep-th/9112038} \BibitemShut
  {NoStop}%
\bibitem [{\citenamefont {Preskill}\ and\ \citenamefont
  {Krauss}(1990)}]{Preskill:1990bm}%
  \BibitemOpen
  \bibfield  {author} {\bibinfo {author} {\bibfnamefont {J.}~\bibnamefont
  {Preskill}}\ and\ \bibinfo {author} {\bibfnamefont {L.~M.}\ \bibnamefont
  {Krauss}},\ }\href {\doibase 10.1016/0550-3213(90)90262-C} {\bibfield
  {journal} {\bibinfo  {journal} {Nucl. Phys. B}\ }\textbf {\bibinfo {volume}
  {341}},\ \bibinfo {pages} {50} (\bibinfo {year} {1990})}\BibitemShut
  {NoStop}%
\bibitem [{\citenamefont {Bucher}\ \emph {et~al.}(1992)\citenamefont {Bucher},
  \citenamefont {Lo},\ and\ \citenamefont {Preskill}}]{Bucher:1992bd}%
  \BibitemOpen
  \bibfield  {author} {\bibinfo {author} {\bibfnamefont {M.}~\bibnamefont
  {Bucher}}, \bibinfo {author} {\bibfnamefont {H.-K.}\ \bibnamefont {Lo}}, \
  and\ \bibinfo {author} {\bibfnamefont {J.}~\bibnamefont {Preskill}},\ }\href
  {\doibase 10.1016/0550-3213(92)90173-9} {\bibfield  {journal} {\bibinfo
  {journal} {Nucl. Phys. B}\ }\textbf {\bibinfo {volume} {386}},\ \bibinfo
  {pages} {3} (\bibinfo {year} {1992})},\ \Eprint
  {http://arxiv.org/abs/hep-th/9112039} {arXiv:hep-th/9112039} \BibitemShut
  {NoStop}%
\bibitem [{\citenamefont {Lo}\ and\ \citenamefont
  {Preskill}(1993)}]{Lo:1993hp}%
  \BibitemOpen
  \bibfield  {author} {\bibinfo {author} {\bibfnamefont {H.-K.}\ \bibnamefont
  {Lo}}\ and\ \bibinfo {author} {\bibfnamefont {J.}~\bibnamefont {Preskill}},\
  }\href {\doibase 10.1103/PhysRevD.48.4821} {\bibfield  {journal} {\bibinfo
  {journal} {Phys. Rev. D}\ }\textbf {\bibinfo {volume} {48}},\ \bibinfo
  {pages} {4821} (\bibinfo {year} {1993})},\ \Eprint
  {http://arxiv.org/abs/hep-th/9306006} {arXiv:hep-th/9306006} \BibitemShut
  {NoStop}%
\bibitem [{\citenamefont {Leonhardt}\ and\ \citenamefont
  {Volovik}(2000)}]{Leonhardt:2000km}%
  \BibitemOpen
  \bibfield  {author} {\bibinfo {author} {\bibfnamefont {U.}~\bibnamefont
  {Leonhardt}}\ and\ \bibinfo {author} {\bibfnamefont {G.~E.}\ \bibnamefont
  {Volovik}},\ }\href {\doibase 10.1134/1.1312008} {\bibfield  {journal}
  {\bibinfo  {journal} {Pisma Zh. Eksp. Teor. Fiz.}\ }\textbf {\bibinfo
  {volume} {72}},\ \bibinfo {pages} {66} (\bibinfo {year} {2000})},\ \Eprint
  {http://arxiv.org/abs/cond-mat/0003428} {arXiv:cond-mat/0003428} \BibitemShut
  {NoStop}%
\bibitem [{\citenamefont {Chatterjee}\ and\ \citenamefont
  {Nitta}(2017{\natexlab{a}})}]{Chatterjee:2017jsi}%
  \BibitemOpen
  \bibfield  {author} {\bibinfo {author} {\bibfnamefont {C.}~\bibnamefont
  {Chatterjee}}\ and\ \bibinfo {author} {\bibfnamefont {M.}~\bibnamefont
  {Nitta}},\ }\href {\doibase 10.1007/JHEP09(2017)046} {\bibfield  {journal}
  {\bibinfo  {journal} {JHEP}\ }\textbf {\bibinfo {volume} {09}},\ \bibinfo
  {pages} {046} (\bibinfo {year} {2017}{\natexlab{a}})},\ \Eprint
  {http://arxiv.org/abs/1703.08971} {arXiv:1703.08971 [hep-th]} \BibitemShut
  {NoStop}%
\bibitem [{\citenamefont {Chatterjee}\ and\ \citenamefont
  {Nitta}(2017{\natexlab{b}})}]{Chatterjee:2017hya}%
  \BibitemOpen
  \bibfield  {author} {\bibinfo {author} {\bibfnamefont {C.}~\bibnamefont
  {Chatterjee}}\ and\ \bibinfo {author} {\bibfnamefont {M.}~\bibnamefont
  {Nitta}},\ }\href {\doibase 10.1140/epjc/s10052-017-5352-1} {\bibfield
  {journal} {\bibinfo  {journal} {Eur. Phys. J. C}\ }\textbf {\bibinfo {volume}
  {77}},\ \bibinfo {pages} {809} (\bibinfo {year} {2017}{\natexlab{b}})},\
  \Eprint {http://arxiv.org/abs/1706.10212} {arXiv:1706.10212 [hep-th]}
  \BibitemShut {NoStop}%
\bibitem [{\citenamefont {Chatterjee}\ and\ \citenamefont
  {Nitta}(2020)}]{Chatterjee:2019zwx}%
  \BibitemOpen
  \bibfield  {author} {\bibinfo {author} {\bibfnamefont {C.}~\bibnamefont
  {Chatterjee}}\ and\ \bibinfo {author} {\bibfnamefont {M.}~\bibnamefont
  {Nitta}},\ }\href {\doibase 10.1103/PhysRevD.101.085002} {\bibfield
  {journal} {\bibinfo  {journal} {Phys. Rev. D}\ }\textbf {\bibinfo {volume}
  {101}},\ \bibinfo {pages} {085002} (\bibinfo {year} {2020})},\ \Eprint
  {http://arxiv.org/abs/1905.01884} {arXiv:1905.01884 [hep-th]} \BibitemShut
  {NoStop}%
\bibitem [{\citenamefont {Cipriani}\ \emph {et~al.}(2012)\citenamefont
  {Cipriani}, \citenamefont {Vinci},\ and\ \citenamefont
  {Nitta}}]{Cipriani:2012hr}%
  \BibitemOpen
  \bibfield  {author} {\bibinfo {author} {\bibfnamefont {M.}~\bibnamefont
  {Cipriani}}, \bibinfo {author} {\bibfnamefont {W.}~\bibnamefont {Vinci}}, \
  and\ \bibinfo {author} {\bibfnamefont {M.}~\bibnamefont {Nitta}},\ }\href
  {\doibase 10.1103/PhysRevD.86.121704} {\bibfield  {journal} {\bibinfo
  {journal} {Phys. Rev. D}\ }\textbf {\bibinfo {volume} {86}},\ \bibinfo
  {pages} {121704} (\bibinfo {year} {2012})},\ \Eprint
  {http://arxiv.org/abs/1208.5704} {arXiv:1208.5704 [hep-ph]} \BibitemShut
  {NoStop}%
\bibitem [{\citenamefont {Alford}\ \emph {et~al.}(2016)\citenamefont {Alford},
  \citenamefont {Mallavarapu}, \citenamefont {Vachaspati},\ and\ \citenamefont
  {Windisch}}]{Alford:2016dco}%
  \BibitemOpen
  \bibfield  {author} {\bibinfo {author} {\bibfnamefont {M.~G.}\ \bibnamefont
  {Alford}}, \bibinfo {author} {\bibfnamefont {S.}~\bibnamefont {Mallavarapu}},
  \bibinfo {author} {\bibfnamefont {T.}~\bibnamefont {Vachaspati}}, \ and\
  \bibinfo {author} {\bibfnamefont {A.}~\bibnamefont {Windisch}},\ }\href
  {\doibase 10.1103/PhysRevC.93.045801} {\bibfield  {journal} {\bibinfo
  {journal} {Phys. Rev. C}\ }\textbf {\bibinfo {volume} {93}},\ \bibinfo
  {pages} {045801} (\bibinfo {year} {2016})},\ \Eprint
  {http://arxiv.org/abs/1601.04656} {arXiv:1601.04656 [nucl-th]} \BibitemShut
  {NoStop}%
\bibitem [{\citenamefont {Eto}\ \emph {et~al.}(2009)\citenamefont {Eto},
  \citenamefont {Nakano},\ and\ \citenamefont {Nitta}}]{Eto:2009bh}%
  \BibitemOpen
  \bibfield  {author} {\bibinfo {author} {\bibfnamefont {M.}~\bibnamefont
  {Eto}}, \bibinfo {author} {\bibfnamefont {E.}~\bibnamefont {Nakano}}, \ and\
  \bibinfo {author} {\bibfnamefont {M.}~\bibnamefont {Nitta}},\ }\href
  {\doibase 10.1103/PhysRevD.80.125011} {\bibfield  {journal} {\bibinfo
  {journal} {Phys. Rev. D}\ }\textbf {\bibinfo {volume} {80}},\ \bibinfo
  {pages} {125011} (\bibinfo {year} {2009})},\ \Eprint
  {http://arxiv.org/abs/0908.4470} {arXiv:0908.4470 [hep-ph]} \BibitemShut
  {NoStop}%
\bibitem [{\citenamefont {Eto}\ \emph {et~al.}(2010)\citenamefont {Eto},
  \citenamefont {Nitta},\ and\ \citenamefont {Yamamoto}}]{Eto:2009tr}%
  \BibitemOpen
  \bibfield  {author} {\bibinfo {author} {\bibfnamefont {M.}~\bibnamefont
  {Eto}}, \bibinfo {author} {\bibfnamefont {M.}~\bibnamefont {Nitta}}, \ and\
  \bibinfo {author} {\bibfnamefont {N.}~\bibnamefont {Yamamoto}},\ }\href
  {\doibase 10.1103/PhysRevLett.104.161601} {\bibfield  {journal} {\bibinfo
  {journal} {Phys. Rev. Lett.}\ }\textbf {\bibinfo {volume} {104}},\ \bibinfo
  {pages} {161601} (\bibinfo {year} {2010})},\ \Eprint
  {http://arxiv.org/abs/0912.1352} {arXiv:0912.1352 [hep-ph]} \BibitemShut
  {NoStop}%
\bibitem [{\citenamefont {Osterwalder}\ and\ \citenamefont
  {Seiler}(1978)}]{Osterwalder:1977pc}%
  \BibitemOpen
  \bibfield  {author} {\bibinfo {author} {\bibfnamefont {K.}~\bibnamefont
  {Osterwalder}}\ and\ \bibinfo {author} {\bibfnamefont {E.}~\bibnamefont
  {Seiler}},\ }\href {\doibase 10.1016/0003-4916(78)90039-8} {\bibfield
  {journal} {\bibinfo  {journal} {Annals Phys.}\ }\textbf {\bibinfo {volume}
  {110}},\ \bibinfo {pages} {440} (\bibinfo {year} {1978})}\BibitemShut
  {NoStop}%
\bibitem [{\citenamefont {Fradkin}\ and\ \citenamefont
  {Shenker}(1979)}]{Fradkin:1978dv}%
  \BibitemOpen
  \bibfield  {author} {\bibinfo {author} {\bibfnamefont {E.~H.}\ \bibnamefont
  {Fradkin}}\ and\ \bibinfo {author} {\bibfnamefont {S.~H.}\ \bibnamefont
  {Shenker}},\ }\href {\doibase 10.1103/PhysRevD.19.3682} {\bibfield  {journal}
  {\bibinfo  {journal} {Phys. Rev. D}\ }\textbf {\bibinfo {volume} {19}},\
  \bibinfo {pages} {3682} (\bibinfo {year} {1979})}\BibitemShut {NoStop}%
\bibitem [{\citenamefont {Banks}\ and\ \citenamefont
  {Rabinovici}(1979)}]{Banks:1979fi}%
  \BibitemOpen
  \bibfield  {author} {\bibinfo {author} {\bibfnamefont {T.}~\bibnamefont
  {Banks}}\ and\ \bibinfo {author} {\bibfnamefont {E.}~\bibnamefont
  {Rabinovici}},\ }\href {\doibase 10.1016/0550-3213(79)90064-6} {\bibfield
  {journal} {\bibinfo  {journal} {Nucl. Phys. B}\ }\textbf {\bibinfo {volume}
  {160}},\ \bibinfo {pages} {349} (\bibinfo {year} {1979})}\BibitemShut
  {NoStop}%
\bibitem [{Note1()}]{Note1}%
  \BibitemOpen
  \bibinfo {note} {It is known in the nematic phase~\cite {Sauls:1978lna} for
  which $\protect \mathrm {diag}(1,r,1-r)$ is implied for the $i,j$ indices,
  with real parameter $r$.}\BibitemShut {Stop}%
\bibitem [{\citenamefont {Auzzi}\ and\ \citenamefont
  {Shifman}(2007)}]{Auzzi:2006ns}%
  \BibitemOpen
  \bibfield  {author} {\bibinfo {author} {\bibfnamefont {R.}~\bibnamefont
  {Auzzi}}\ and\ \bibinfo {author} {\bibfnamefont {M.}~\bibnamefont
  {Shifman}},\ }\href {\doibase 10.1088/1751-8113/40/23/015} {\bibfield
  {journal} {\bibinfo  {journal} {J. Phys. A}\ }\textbf {\bibinfo {volume}
  {40}},\ \bibinfo {pages} {6221} (\bibinfo {year} {2007})},\ \Eprint
  {http://arxiv.org/abs/hep-th/0612211} {arXiv:hep-th/0612211} \BibitemShut
  {NoStop}%
\bibitem [{\citenamefont {Auzzi}\ \emph {et~al.}(2008)\citenamefont {Auzzi},
  \citenamefont {Bolognesi},\ and\ \citenamefont {Shifman}}]{Auzzi:2008hu}%
  \BibitemOpen
  \bibfield  {author} {\bibinfo {author} {\bibfnamefont {R.}~\bibnamefont
  {Auzzi}}, \bibinfo {author} {\bibfnamefont {S.}~\bibnamefont {Bolognesi}}, \
  and\ \bibinfo {author} {\bibfnamefont {M.}~\bibnamefont {Shifman}},\ }\href
  {\doibase 10.1103/PhysRevD.77.125029} {\bibfield  {journal} {\bibinfo
  {journal} {Phys. Rev. D}\ }\textbf {\bibinfo {volume} {77}},\ \bibinfo
  {pages} {125029} (\bibinfo {year} {2008})},\ \Eprint
  {http://arxiv.org/abs/0804.0229} {arXiv:0804.0229 [hep-th]} \BibitemShut
  {NoStop}%
\bibitem [{\citenamefont {Iida}(2005)}]{Iida:2004if}%
  \BibitemOpen
  \bibfield  {author} {\bibinfo {author} {\bibfnamefont {K.}~\bibnamefont
  {Iida}},\ }\href {\doibase 10.1103/PhysRevD.71.054011} {\bibfield  {journal}
  {\bibinfo  {journal} {Phys. Rev. D}\ }\textbf {\bibinfo {volume} {71}},\
  \bibinfo {pages} {054011} (\bibinfo {year} {2005})},\ \Eprint
  {http://arxiv.org/abs/hep-ph/0412426} {arXiv:hep-ph/0412426} \BibitemShut
  {NoStop}%
\bibitem [{Note2()}]{Note2}%
  \BibitemOpen
  \bibinfo {note} {This is in contrast to the case of a non-Abelian vortex in
  the CFL phase, accompanied by the $\protect \mathbb {C} P^2$ orientational
  moduli \cite {Nakano:2007dr,Eto:2009bh,Eto:2013hoa,Eto:2009tr}.}\BibitemShut
  {Stop}%
\bibitem [{Note3()}]{Note3}%
  \BibitemOpen
  \bibinfo {note} {In Refs.~\cite
  {Chatterjee:2017jsi,Chatterjee:2017hya,Chatterjee:2019zwx}, an SU(2) $\times
  $ U(1) gauge theory with charged triplet scalar fields were studied, which is
  an SU(2) version of our case where U(1) is also gauged. The case of a global
  U(1) symmetry, closer to our case, was discussed in Refs.~\cite
  {Sato:2018nqy,Chatterjee:2019rch} in the context of the axion dark matter
  model.}\BibitemShut {Stop}%
\bibitem [{\citenamefont {Chatterjee}\ and\ \citenamefont
  {Nitta}(2016)}]{Chatterjee:2015lbf}%
  \BibitemOpen
  \bibfield  {author} {\bibinfo {author} {\bibfnamefont {C.}~\bibnamefont
  {Chatterjee}}\ and\ \bibinfo {author} {\bibfnamefont {M.}~\bibnamefont
  {Nitta}},\ }\href {\doibase 10.1103/PhysRevD.93.065050} {\bibfield  {journal}
  {\bibinfo  {journal} {Phys. Rev. D}\ }\textbf {\bibinfo {volume} {93}},\
  \bibinfo {pages} {065050} (\bibinfo {year} {2016})},\ \Eprint
  {http://arxiv.org/abs/1512.06603} {arXiv:1512.06603 [hep-ph]} \BibitemShut
  {NoStop}%
\bibitem [{Note4()}]{Note4}%
  \BibitemOpen
  \bibinfo {note} {This is in contrast to the case of non-Abelian vortices
  (color flux tubes) in the CFL phase, around which all (generalized) AB phases
  are color singlet \cite {Chatterjee:2018nxe,
  *Chatterjee:2019tbz}.}\BibitemShut {Stop}%
\bibitem [{\citenamefont {Vinci}\ \emph {et~al.}(2012)\citenamefont {Vinci},
  \citenamefont {Cipriani},\ and\ \citenamefont {Nitta}}]{Vinci:2012mc}%
  \BibitemOpen
  \bibfield  {author} {\bibinfo {author} {\bibfnamefont {W.}~\bibnamefont
  {Vinci}}, \bibinfo {author} {\bibfnamefont {M.}~\bibnamefont {Cipriani}}, \
  and\ \bibinfo {author} {\bibfnamefont {M.}~\bibnamefont {Nitta}},\ }\href
  {\doibase 10.1103/PhysRevD.86.085018} {\bibfield  {journal} {\bibinfo
  {journal} {Phys. Rev. D}\ }\textbf {\bibinfo {volume} {86}},\ \bibinfo
  {pages} {085018} (\bibinfo {year} {2012})},\ \Eprint
  {http://arxiv.org/abs/1206.3535} {arXiv:1206.3535 [hep-ph]} \BibitemShut
  {NoStop}%
\bibitem [{\citenamefont {Masuda}\ and\ \citenamefont
  {Nitta}(2020)}]{Masuda:2016vak}%
  \BibitemOpen
  \bibfield  {author} {\bibinfo {author} {\bibfnamefont {K.}~\bibnamefont
  {Masuda}}\ and\ \bibinfo {author} {\bibfnamefont {M.}~\bibnamefont {Nitta}},\
  }\href {\doibase 10.1093/ptep/ptz138} {\bibfield  {journal} {\bibinfo
  {journal} {PTEP}\ }\textbf {\bibinfo {volume} {2020}},\ \bibinfo {pages}
  {013D01} (\bibinfo {year} {2020})},\ \Eprint
  {http://arxiv.org/abs/1602.07050} {arXiv:1602.07050 [nucl-th]} \BibitemShut
  {NoStop}%
\bibitem [{\citenamefont {Muzikar}\ \emph {et~al.}(1980)\citenamefont
  {Muzikar}, \citenamefont {Sauls},\ and\ \citenamefont
  {Serene}}]{Muzikar:1980as}%
  \BibitemOpen
  \bibfield  {author} {\bibinfo {author} {\bibfnamefont {P.}~\bibnamefont
  {Muzikar}}, \bibinfo {author} {\bibfnamefont {J.}~\bibnamefont {Sauls}}, \
  and\ \bibinfo {author} {\bibfnamefont {J.}~\bibnamefont {Serene}},\ }\href
  {\doibase 10.1103/PhysRevD.21.1494} {\bibfield  {journal} {\bibinfo
  {journal} {Phys. Rev. D}\ }\textbf {\bibinfo {volume} {21}},\ \bibinfo
  {pages} {1494} (\bibinfo {year} {1980})}\BibitemShut {NoStop}%
\bibitem [{\citenamefont {Sauls}\ \emph {et~al.}(1982)\citenamefont {Sauls},
  \citenamefont {Stein},\ and\ \citenamefont {Serene}}]{Sauls:1982ie}%
  \BibitemOpen
  \bibfield  {author} {\bibinfo {author} {\bibfnamefont {J.}~\bibnamefont
  {Sauls}}, \bibinfo {author} {\bibfnamefont {D.}~\bibnamefont {Stein}}, \ and\
  \bibinfo {author} {\bibfnamefont {J.}~\bibnamefont {Serene}},\ }\href
  {\doibase 10.1103/PhysRevD.25.967} {\bibfield  {journal} {\bibinfo  {journal}
  {Phys. Rev. D}\ }\textbf {\bibinfo {volume} {25}},\ \bibinfo {pages} {967}
  (\bibinfo {year} {1982})}\BibitemShut {NoStop}%
\bibitem [{\citenamefont {Masuda}\ and\ \citenamefont
  {Nitta}(2016)}]{Masuda:2015jka}%
  \BibitemOpen
  \bibfield  {author} {\bibinfo {author} {\bibfnamefont {K.}~\bibnamefont
  {Masuda}}\ and\ \bibinfo {author} {\bibfnamefont {M.}~\bibnamefont {Nitta}},\
  }\href {\doibase 10.1103/PhysRevC.93.035804} {\bibfield  {journal} {\bibinfo
  {journal} {Phys. Rev. C}\ }\textbf {\bibinfo {volume} {93}},\ \bibinfo
  {pages} {035804} (\bibinfo {year} {2016})},\ \Eprint
  {http://arxiv.org/abs/1512.01946} {arXiv:1512.01946 [nucl-th]} \BibitemShut
  {NoStop}%
\bibitem [{\citenamefont {Chatterjee}\ \emph {et~al.}(2017)\citenamefont
  {Chatterjee}, \citenamefont {Haberichter},\ and\ \citenamefont
  {Nitta}}]{Chatterjee:2016gpm}%
  \BibitemOpen
  \bibfield  {author} {\bibinfo {author} {\bibfnamefont {C.}~\bibnamefont
  {Chatterjee}}, \bibinfo {author} {\bibfnamefont {M.}~\bibnamefont
  {Haberichter}}, \ and\ \bibinfo {author} {\bibfnamefont {M.}~\bibnamefont
  {Nitta}},\ }\href {\doibase 10.1103/PhysRevC.96.055807} {\bibfield  {journal}
  {\bibinfo  {journal} {Phys. Rev. C}\ }\textbf {\bibinfo {volume} {96}},\
  \bibinfo {pages} {055807} (\bibinfo {year} {2017})},\ \Eprint
  {http://arxiv.org/abs/1612.05588} {arXiv:1612.05588 [nucl-th]} \BibitemShut
  {NoStop}%
\bibitem [{\citenamefont {Masaki}\ \emph {et~al.}(2020)\citenamefont {Masaki},
  \citenamefont {Mizushima},\ and\ \citenamefont {Nitta}}]{Masaki:2019rsz}%
  \BibitemOpen
  \bibfield  {author} {\bibinfo {author} {\bibfnamefont {Y.}~\bibnamefont
  {Masaki}}, \bibinfo {author} {\bibfnamefont {T.}~\bibnamefont {Mizushima}}, \
  and\ \bibinfo {author} {\bibfnamefont {M.}~\bibnamefont {Nitta}},\ }\href
  {\doibase 10.1103/PhysRevResearch.2.013193} {\bibfield  {journal} {\bibinfo
  {journal} {Phys. Rev. Res.}\ }\textbf {\bibinfo {volume} {2}},\ \bibinfo
  {pages} {013193} (\bibinfo {year} {2020})},\ \Eprint
  {http://arxiv.org/abs/1908.06215} {arXiv:1908.06215 [cond-mat.supr-con]}
  \BibitemShut {NoStop}%
\bibitem [{\citenamefont {Yasui}\ \emph {et~al.}(2010)\citenamefont {Yasui},
  \citenamefont {Itakura},\ and\ \citenamefont {Nitta}}]{Yasui:2010yw}%
  \BibitemOpen
  \bibfield  {author} {\bibinfo {author} {\bibfnamefont {S.}~\bibnamefont
  {Yasui}}, \bibinfo {author} {\bibfnamefont {K.}~\bibnamefont {Itakura}}, \
  and\ \bibinfo {author} {\bibfnamefont {M.}~\bibnamefont {Nitta}},\ }\href
  {\doibase 10.1103/PhysRevD.81.105003} {\bibfield  {journal} {\bibinfo
  {journal} {Phys. Rev. D}\ }\textbf {\bibinfo {volume} {81}},\ \bibinfo
  {pages} {105003} (\bibinfo {year} {2010})},\ \Eprint
  {http://arxiv.org/abs/1001.3730} {arXiv:1001.3730 [math-ph]} \BibitemShut
  {NoStop}%
\bibitem [{\citenamefont {Fujiwara}\ \emph {et~al.}(2011)\citenamefont
  {Fujiwara}, \citenamefont {Fukui}, \citenamefont {Nitta},\ and\ \citenamefont
  {Yasui}}]{Fujiwara:2011za}%
  \BibitemOpen
  \bibfield  {author} {\bibinfo {author} {\bibfnamefont {T.}~\bibnamefont
  {Fujiwara}}, \bibinfo {author} {\bibfnamefont {T.}~\bibnamefont {Fukui}},
  \bibinfo {author} {\bibfnamefont {M.}~\bibnamefont {Nitta}}, \ and\ \bibinfo
  {author} {\bibfnamefont {S.}~\bibnamefont {Yasui}},\ }\href {\doibase
  10.1103/PhysRevD.84.076002} {\bibfield  {journal} {\bibinfo  {journal} {Phys.
  Rev. D}\ }\textbf {\bibinfo {volume} {84}},\ \bibinfo {pages} {076002}
  (\bibinfo {year} {2011})},\ \Eprint {http://arxiv.org/abs/1105.2115}
  {arXiv:1105.2115 [hep-ph]} \BibitemShut {NoStop}%
\bibitem [{\citenamefont {Hidaka}\ \emph {et~al.}(2019)\citenamefont {Hidaka},
  \citenamefont {Hirono}, \citenamefont {Nitta}, \citenamefont {Tanizaki},\
  and\ \citenamefont {Yokokura}}]{Hidaka:2019jtv}%
  \BibitemOpen
  \bibfield  {author} {\bibinfo {author} {\bibfnamefont {Y.}~\bibnamefont
  {Hidaka}}, \bibinfo {author} {\bibfnamefont {Y.}~\bibnamefont {Hirono}},
  \bibinfo {author} {\bibfnamefont {M.}~\bibnamefont {Nitta}}, \bibinfo
  {author} {\bibfnamefont {Y.}~\bibnamefont {Tanizaki}}, \ and\ \bibinfo
  {author} {\bibfnamefont {R.}~\bibnamefont {Yokokura}},\ }\href {\doibase
  10.1103/PhysRevD.100.125016} {\bibfield  {journal} {\bibinfo  {journal}
  {Phys. Rev. D}\ }\textbf {\bibinfo {volume} {100}},\ \bibinfo {pages}
  {125016} (\bibinfo {year} {2019})},\ \Eprint
  {http://arxiv.org/abs/1903.06389} {arXiv:1903.06389 [hep-th]} \BibitemShut
  {NoStop}%
\bibitem [{\citenamefont {Ivanov}(2001)}]{Ivanov:2000mjr}%
  \BibitemOpen
  \bibfield  {author} {\bibinfo {author} {\bibfnamefont {D.}~\bibnamefont
  {Ivanov}},\ }\href {\doibase 10.1103/PhysRevLett.86.268} {\bibfield
  {journal} {\bibinfo  {journal} {Phys. Rev. Lett.}\ }\textbf {\bibinfo
  {volume} {86}},\ \bibinfo {pages} {268} (\bibinfo {year} {2001})},\ \Eprint
  {http://arxiv.org/abs/cond-mat/0005069} {arXiv:cond-mat/0005069} \BibitemShut
  {NoStop}%
\bibitem [{\citenamefont {Yasui}\ \emph {et~al.}(2011)\citenamefont {Yasui},
  \citenamefont {Itakura},\ and\ \citenamefont {Nitta}}]{Yasui:2010yh}%
  \BibitemOpen
  \bibfield  {author} {\bibinfo {author} {\bibfnamefont {S.}~\bibnamefont
  {Yasui}}, \bibinfo {author} {\bibfnamefont {K.}~\bibnamefont {Itakura}}, \
  and\ \bibinfo {author} {\bibfnamefont {M.}~\bibnamefont {Nitta}},\ }\href
  {\doibase 10.1103/PhysRevB.83.134518} {\bibfield  {journal} {\bibinfo
  {journal} {Phys. Rev. B}\ }\textbf {\bibinfo {volume} {83}},\ \bibinfo
  {pages} {134518} (\bibinfo {year} {2011})},\ \Eprint
  {http://arxiv.org/abs/1010.3331} {arXiv:1010.3331 [cond-mat.mes-hall]}
  \BibitemShut {NoStop}%
\bibitem [{\citenamefont {Hirono}\ \emph {et~al.}(2012)\citenamefont {Hirono},
  \citenamefont {Yasui}, \citenamefont {Itakura},\ and\ \citenamefont
  {Nitta}}]{Hirono:2012ad}%
  \BibitemOpen
  \bibfield  {author} {\bibinfo {author} {\bibfnamefont {Y.}~\bibnamefont
  {Hirono}}, \bibinfo {author} {\bibfnamefont {S.}~\bibnamefont {Yasui}},
  \bibinfo {author} {\bibfnamefont {K.}~\bibnamefont {Itakura}}, \ and\
  \bibinfo {author} {\bibfnamefont {M.}~\bibnamefont {Nitta}},\ }\href
  {\doibase 10.1103/PhysRevB.86.014508} {\bibfield  {journal} {\bibinfo
  {journal} {Phys. Rev. B}\ }\textbf {\bibinfo {volume} {86}},\ \bibinfo
  {pages} {014508} (\bibinfo {year} {2012})},\ \Eprint
  {http://arxiv.org/abs/1203.0173} {arXiv:1203.0173 [cond-mat.supr-con]}
  \BibitemShut {NoStop}%
\bibitem [{\citenamefont {Sato}\ \emph {et~al.}(2018)\citenamefont {Sato},
  \citenamefont {Takahashi},\ and\ \citenamefont {Yamada}}]{Sato:2018nqy}%
  \BibitemOpen
  \bibfield  {author} {\bibinfo {author} {\bibfnamefont {R.}~\bibnamefont
  {Sato}}, \bibinfo {author} {\bibfnamefont {F.}~\bibnamefont {Takahashi}}, \
  and\ \bibinfo {author} {\bibfnamefont {M.}~\bibnamefont {Yamada}},\ }\href
  {\doibase 10.1103/PhysRevD.98.043535} {\bibfield  {journal} {\bibinfo
  {journal} {Phys. Rev. D}\ }\textbf {\bibinfo {volume} {98}},\ \bibinfo
  {pages} {043535} (\bibinfo {year} {2018})},\ \Eprint
  {http://arxiv.org/abs/1805.10533} {arXiv:1805.10533 [hep-ph]} \BibitemShut
  {NoStop}%
\bibitem [{\citenamefont {Chatterjee}\ \emph {et~al.}(2020)\citenamefont
  {Chatterjee}, \citenamefont {Higaki},\ and\ \citenamefont
  {Nitta}}]{Chatterjee:2019rch}%
  \BibitemOpen
  \bibfield  {author} {\bibinfo {author} {\bibfnamefont {C.}~\bibnamefont
  {Chatterjee}}, \bibinfo {author} {\bibfnamefont {T.}~\bibnamefont {Higaki}},
  \ and\ \bibinfo {author} {\bibfnamefont {M.}~\bibnamefont {Nitta}},\ }\href
  {\doibase 10.1103/PhysRevD.101.075026} {\bibfield  {journal} {\bibinfo
  {journal} {Phys. Rev. D}\ }\textbf {\bibinfo {volume} {101}},\ \bibinfo
  {pages} {075026} (\bibinfo {year} {2020})},\ \Eprint
  {http://arxiv.org/abs/1903.11753} {arXiv:1903.11753 [hep-ph]} \BibitemShut
  {NoStop}%
\end{thebibliography}%

\end{document}